\documentclass[aip,reprint]{revtex4-2}

\usepackage{graphicx,color}

\newcommand{\bfr}{{\bf r}}
\newcommand{\bfR}{{\bf R}}
\newcommand{\bfq}{{\bf q}}
\newcommand{\bfF}{{\bf F}}

\newcommand{\bfj}{{\bf j}}
\newcommand{\bfA}{{\bf A}}
\newcommand{\ua}{\uparrow}
\newcommand{\da}{\downarrow}

\begin{document}

\title{A Snapshot of Time-Dependent Density-Functional Theory}
\author{Carsten A. Ullrich}
\affiliation{Department of Physics and Astronomy, University of Missouri, Columbia, Missouri 65211, USA}

\begin{abstract}
Time-dependent density-functional theory (TDDFT) is an extension of ground-state density-functional theory which allows the treatment
of electronic excited states and a wide range of time-dependent phenomena in the linear and nonlinear regime, including coupled electron-nuclear dynamics.
TDDFT is a vibrant field with many exciting applications in physics, (bio)chemistry, materials science and other areas. This perspective
gives an overview of recent developments and successes, formal and computational challenges, and hot topics in TDDFT.
\end{abstract}

\maketitle

\section{Introduction}

Time-dependent density-functional theory (TDDFT) was formally established in 1984 in the seminal work of Runge and Gross,\cite{Runge1984}
20 years after Hohenberg, Kohn and Sham had laid the foundations of ground-state density-functional theory (DFT).\cite{Hohenberg1964,Kohn1965}
Over the past decades, DFT has become the dominant method to describe the electronic and structural properties of materials from first principles,
due to its still unsurpassed combination of useful accuracy and computational efficiency. The range of applications of DFT
is vast, and so is the sheer number and diversity of the scientists who are using DFT in practice. The successes and challenges of DFT
have been widely reviewed\cite{Kohn1999,Burke2012,Perdew2013,Becke2014,Jones2015,Kaplan2023} and have become textbook material.\cite{Giustino2014,Martin2020}

\begin{figure}
  \includegraphics[width=\linewidth]{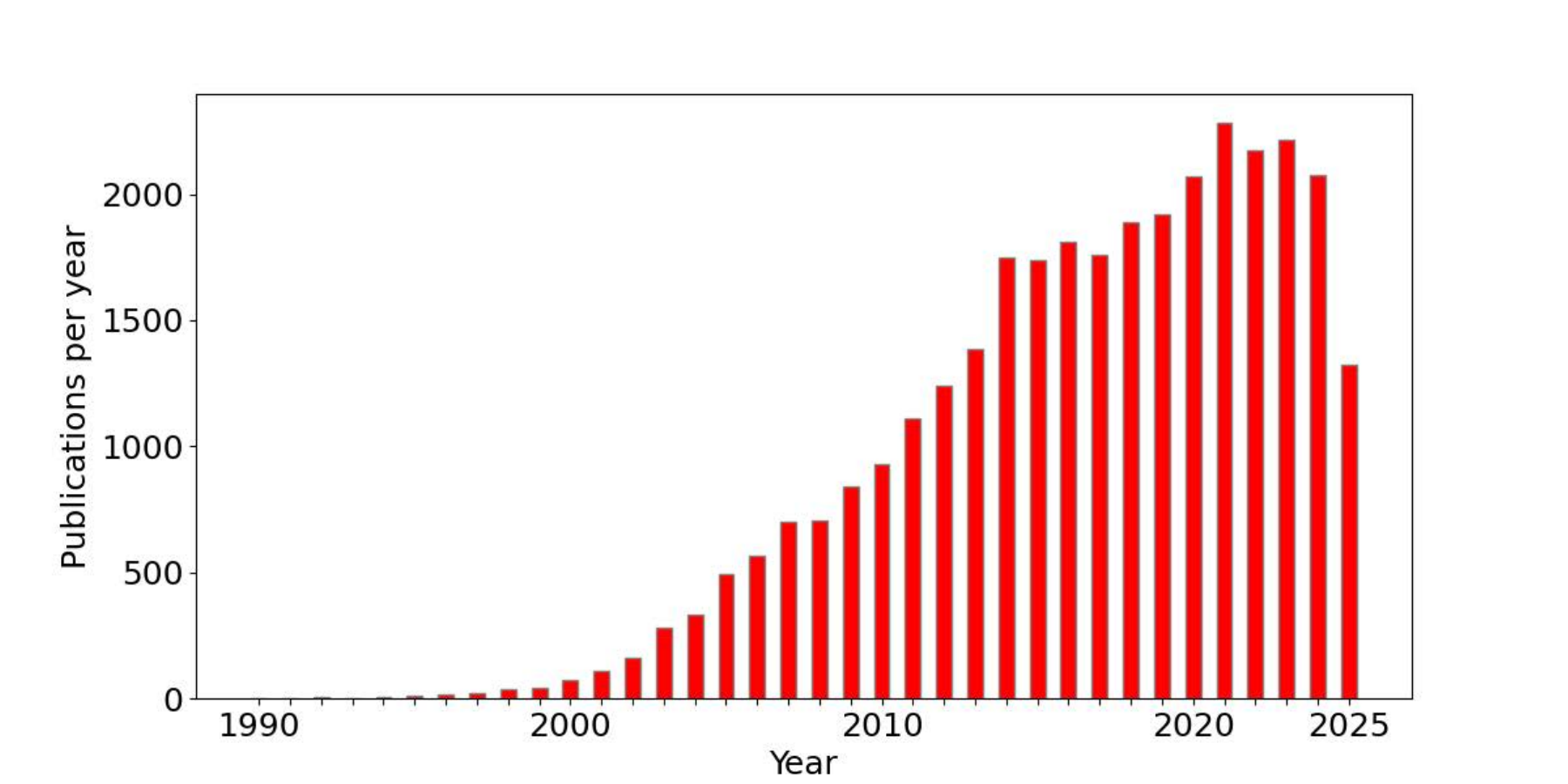}
  \caption{\label{fig1} Number of papers per year using or citing TDDFT, from Web of Science (accessed on September 12, 2025).\cite{WOS}}
\end{figure}

In turn, TDDFT has been the subject of a textbook \cite{Ullrich2012}, two edited volumes,\cite{TDDFT2006,TDDFT2012}
and several recent review articles.\cite{Burke2005,Casida2009,Ullrich2014,Maitra2016,Provorse2016,Goings2018,Li2020,You2020,Herbert2023,Xu2024,Sato2025}
A practical measure of the popularity of a research field is the number of publications it generates per year. For TDDFT, we see from Fig. \ref{fig1}
that this number has strongly increased over the past decades, with roughly 2000 papers per year at present. Compared to DFT, which has an order of magnitude more publications
per year,\cite{Jones2015} this seems relatively small. However, the field of TDDFT has a vibrant and active worldwide community of developers and users,
with a very high level of innovation and an ever increasing scope of applications. It is, without doubt, a fast moving area of research,
and in this perspective an attempt is made to provide a snapshot to see where the field is currently standing, and to make reasonable conjectures in which directions it is heading.

Needless to say, this perspective is not intended as a comprehensive review. It is addressed at an audience assumed to be familiar with the basics of electronic structure
theory and DFT, and curious about the current status and hot topics of TDDFT. The majority of applications of TDDFT have been, and still are, in the calculation
of excitation energies and optical spectra of molecular systems, but there have been many exciting recent developments in condensed-matter physics and materials science,
made possible by a proliferation of computer codes and software for TDDFT in the linear-response (LR) and real-time (RT) regime. We will discuss these topics, focusing
on recent developments of TDDFT within the last 10 years.

\section{Fundamentals}\label{sec:2}

\subsection{Existence proofs}

Density-functional theories are notorious for the formal and mathematical subtleties associated with rigorous existence proofs.
In ground-state DFT, an extensive body of work has been dedicated to analyzing its foundations, which was comprehensively reviewed by Penz {\em et al.}\cite{Penz2023a,Penz2023b}
TDDFT poses additional challenges not present in ground-state theories and requires methods of proof that are quite different from DFT, for two main reasons:
\begin{enumerate}
\item Unlike ground-state DFT, TDDFT is not based on a variational minimum principle. TDDFT stationary action principles have
been formulated\cite{VanLeeuwen1998,Vignale2008} but not been explicitly invoked in any attempt to establish rigorous mathematical existence proofs.

\item TDDFT represents an initial value problem, where a given initial state is propagated forward in time.
In the time-dependent Kohn-Sham (TDKS) scheme (see below), the Hamiltonian depends on the time-evolving density,
which implies a different type of self-consistency than in ground-state DFT; time-propagation algorithms must account for this.\cite{Ullrich2012}

\end{enumerate}

The original formulation of TDDFT was developed for systems of $N$ electrons which obey the nonrelativistic
time-dependent Schr\"odinger equation
\begin{equation}\label{SE}
i\frac{\partial}{\partial t} \Psi(\bfr_1,\ldots,\bfr_N,t) = \hat H(t) \Psi(\bfr_1,\ldots,\bfr_N,t) ,
\end{equation}
with the many-body Hamiltonian
\begin{equation}
\hat H(t) = \sum_{j=1}^N \left(-\frac{\nabla_j^2}{2} + v(\bfr_j,t)\right) + \frac{1}{2} \sum_{j\ne l}^N \frac{1}{|\bfr_j - \bfr_l|} \:.
\end{equation}
Here and in the following, atomic units are used and, at least for now, we ignore the spin of the electrons.

Runge and Gross showed \cite{Runge1984} that there is a one-to-one correspondence between the time-dependent scalar potential $v(\bfr,t)$
and the probability density $n(\bfr,t)$ (which, like in ground-state DFT, follows from $|\Psi|^2$ by integrating over all coordinates $\bfr_j$ except one).
Thus, for a given initial many-body state $\Psi_0$, the densities $n(\bfr,t)$ and $n'(\bfr,t)$ associated with
two potentials $v(\bfr,t)$ and $v'(\bfr,t)$ [where $v'(\bfr,t) \ne v(\bfr,t)+ c(t)$]
will always be different for $t>t_0$. Hence, $\Psi$ and everything that can be known from it is uniquely determined by $n(\bfr,t)$ and $\Psi_0$.
The proof proceeds in two steps: first, it is shown that different potentials always generate different current densities $\bfj(\bfr,t)$,
and then the continuity equation implies that this also leads to different densities.
The proof involves an expansion of $v(\bfr,t)$ in a Taylor series about the initial time $t_0$.

In practice, $n(\bfr,t)$ is obtained from the TDKS equation
\begin{equation}\label{TDKS}
i \frac{\partial}{\partial t} \varphi_j(\bfr,t) = \bigg( -\frac{\nabla^2}{2} + v(\bfr,t)
+v_{\rm H}(\bfr,t)  +v_{\rm xc}(\bfr,t)\bigg)\varphi_j(\bfr,t)\:,
\end{equation}
where $v_{\rm H}(\bfr,t) = \int d\bfr' n(\bfr',t)/|\bfr - \bfr'| $ and  $v_{\rm xc}(\bfr,t)$ are the time-dependent Hartree and exchange-correlation (xc) potentials, respectively. The time-dependent density then follows from the Kohn-Sham orbitals as
\begin{equation}\label{TDKS_n}
n(\bfr,t) = \sum_{j=1}^N |\varphi_j(\bfr,t)|^2 \:,
\end{equation}
and whatever one is interested in, such as the time-dependent dipole moment or any other observable of the system, can then in principle be obtained from $n(\bfr,t)$.
Some quantities are not easy to write as explicit functionals of the density (for instance, scattering cross sections in collision processes\cite{Kirchner2024}), and
additional approximations are required. In such cases it is often straightforward to find approximate expressions in terms of the TDKS orbitals; an example is the
time-dependent exciton wave function discussed in Sec. \ref{sec:RTexcitons}, see Eq. (\ref{TDTDM}). 

To establish the TDKS scheme, Runge and Gross assumed noninteracting $v$-representability, i.e., for any given density $n(\bfr,t)$ of the many-body system there must be
a noninteracting system, with potential $v(\bfr,t) + v_{\rm H}(\bfr,t) + v_{\rm xc}(\bfr,t)$, that reproduces it.
The so-called noninteracting $v$-representability problem was solved in 1999 by van Leeuwen, \cite{VanLeeuwen1999} who proposed an alternative
way to construct the time-dependent potential for a given density, using the equation of motion
\begin{equation}\label{EOM}
\frac{\partial^2n(\bfr,t)}{\partial t^2}  = \nabla \cdot \left[n(\bfr,t) \nabla v(\bfr,t) + \bfF^{\rm kin}(\bfr,t) + \bfF^{\rm int}(\bfr,t) \right],
\end{equation}
where $\bfF^{\rm kin}$ and $\bfF^{\rm int}$ are the internal force densities of the many-body system due to kinetic and interaction effects;
the $\bfF$'s are defined in Ref. \onlinecite{VanLeeuwen1999} and can be expressed
in terms of the one- and two-body reduced density matrix of the many-body system,\cite{Ullrich2012} see also Sec. \ref{sec:IID3} below.
We note that the Runge-Gross and van Leeuwen proofs can also be formulated in the framework of generalized Kohn-Sham theory.\cite{Baer2018}
We also mention some numerical studies of the $v$-representability and invertibility of the time-dependent density-potential map.\cite{Jensen2016,Brown2020}

In addition to the Taylor-expandability of $v(\bfr,t)$, van Leeuwen's proof requires the density $n(\bfr,t)$ to be analytic in time at $t_0$.
This seems to be a reasonable condition, so it came as a bit of a  shock to the TDDFT community when it was realized \cite{Maitra2010,Yang2012,Yang2013} that
there are, in fact, large classes of densities that can exhibit a nonanalytic time-dependence -- for instance, densities that have cusps, like in all atoms,
molecules and solids!

This immediately triggered efforts to find existence proofs of TDDFT which do not rely on Taylor-expandability.\cite{Ruggenthaler2011,Ruggenthaler2012,Ruggenthaler2015}
The idea was to construct a fixed-point proof, starting from Eq. (\ref{EOM}), which directly maps the time-dependent density to an external potential
and a given initial state. The conclusion was that such a fixed-point proof could indeed
be established, under some relatively mild assumptions on the differentiability of $\Psi_0$ and $n(\bfr,t)$. The allowed potentials are rather general,
except they cannot be of the Coulomb type associated with point charges; however, rounded-off Coulomb potentials associated with finite-size nuclei are allowed.

This is still the current status of fundamental existence proofs in TDDFT. Over the past 10 years, progress in the rigorous mathematical underpinnings
of TDDFT has somewhat slowed down, despite many open issues.\cite{Fournais2016,Wrighton2023} Some studies are focused on the existence and uniqueness of solutions of the TDKS equation.\cite{Sprengel2017,Ciaramella2021}
Other formal work uses the equation of motion (\ref{EOM}) for a direct definition of xc potentials.\cite{Tchenkoue2019}
It was also suggested\cite{Tarantino2021} to reformulate TDDFT by using the second time derivative of the density, $\partial ^2 n(\bfr,t)/\partial t^2$, as basic variable
instead of the density itself, which leads to a TDKS scheme whose causal structure is more transparent than the original TDKS formalism; this may
provide an alternative starting point for rigorous mathematical treatments.

Other fundamental developments of TDDFT are in terms of generalizations of the formal framework. Perhaps the most significant one  to emerge
in the past few years is for the coupling between electrons and quantized photon fields, which will be discussed in Sec. \ref{sec:4D}.

Another interesting fundamental development is TDDFT for lattice systems, often with an emphasis on strong correlations.\cite{Verdozzi2008,Verdozzi2011,Karlsson2011,Bostrom2019}
Due to the discrete nature of lattices, existence proofs are more easily achievable than
in the continuum case, using methods of linear algebra and the theory of ordinary differential equations.\cite{Farzanehpour2012,Farzanehpour2014}
The $v$-representability problem in lattice-TDDFT has its own interesting aspects:\cite{Li2008,Rossler2018,Brown2020}
on a lattice, a density cannot change arbitrarily fast because the required current flow is limited by the discretized continuity equation.
Recently, these issues were further clarified using an elegant geometrical approach,\cite{Penz2021,Penz2024}
which offers interesting opportunities and links between TDDFT and various mathematical branches such as topology, group theory, and graph theory.

\subsection{Functionals}

The time-dependent xc potential $v_{\rm xc}(\bfr,t)[n,\Psi_0,\Phi_0]$ is formally a functional of all densities $n(\bfr',t')$ with
$t' \le t$, the initial many-body wave function $\Psi_0$, and the initial Kohn-Sham wave function $\Phi_0$ (usually, but not necessarily, a single Slater determinant). If the exact $v_{\rm xc}[n,\Psi_0,\Phi_0](\bfr,t)$ is known, then Eq. (\ref{TDKS}) gives the same density as the
many-body Schr\"odinger equation (\ref{SE}).
If the initial state is the ground state of the system, then the dependence on $\Psi_0$ and $\Phi_0$ goes away, but $v_{\rm xc}[n](\bfr,t)$ still has a memory
in its density dependence.

The standard approach in RT-TDDFT is the adiabatic approximation, where the memory dependence of the xc functional is not taken into account and
one uses instead
\begin{equation}\label{vxc_adia}
v_{\rm xc}^{\rm ad}[n](\bfr,t) =\left. v_{\rm xc}^{\rm gs}[\tilde n](\bfr)\right|_{\tilde n(\bfr) \to n(\bfr,t)} \:.
\end{equation}
In other words, $v_{\rm xc}^{\rm gs}[n_0](\bfr)$, the xc potential of ground-state DFT, is evaluated with the instantaneous time-dependent density at time $t$.
The adiabatic approximation in LR-TDDFT is the frequency-independent xc kernel (see Sec. \ref{sec:LR})
\begin{equation}\label{fxc_adia}
f_{\rm xc}^{\rm ad}[n_0](\bfr,\bfr') = \left.\frac{\delta v_{\rm xc}^{\rm gs}[\tilde n](\bfr)}{\delta \tilde n(\bfr')} \right|_{\tilde n(\bfr) \to n_0(\bfr)},
\end{equation}
where $n_0(\bfr)$ is the ground-state density of the system.

In practice, $v_{\rm xc}^{\rm gs}$ can be any of the many available approximations of ground-state DFT,
so this aspect of functional development is not specific to TDDFT but directly linked to whatever new functionals are invented in DFT.
Generalized Kohn-Sham theory (featuring hybrid functionals and other functionals where the density dependence is implicit) is treated similarly within
the adiabatic approximation.

Going beyond the adiabatic approximation has been a key challenge for TDDFT since its very inception.
There are many circumstances where memory is important:
\begin{itemize}
\item In LR-TDDFT, nonadiabatic xc kernels are required to capture excitations that are not present in the Kohn-Sham spectrum,
most notably double or multiple excitations.\cite{Maitra2022} Correspondingly, to describe such excitations in RT-TDDFT,
the xc potential must have a memory.\cite{Kapoor2016}

\item In extended systems, nonadiabatic effects are important contributors to the lifetimes of quasiparticles such as plasmons, magnons or excitons.

\item More recently, it has been recognized that memory effects are essential whenever a significant transfer of population or of density
occurs on fast time scales.\cite{Lacombe2023}

\end{itemize}

\subsection{Example: Rabi oscillations and adiabatic approximation}\label{sec:Rabi}
\begin{figure}
  \includegraphics[width=\linewidth]{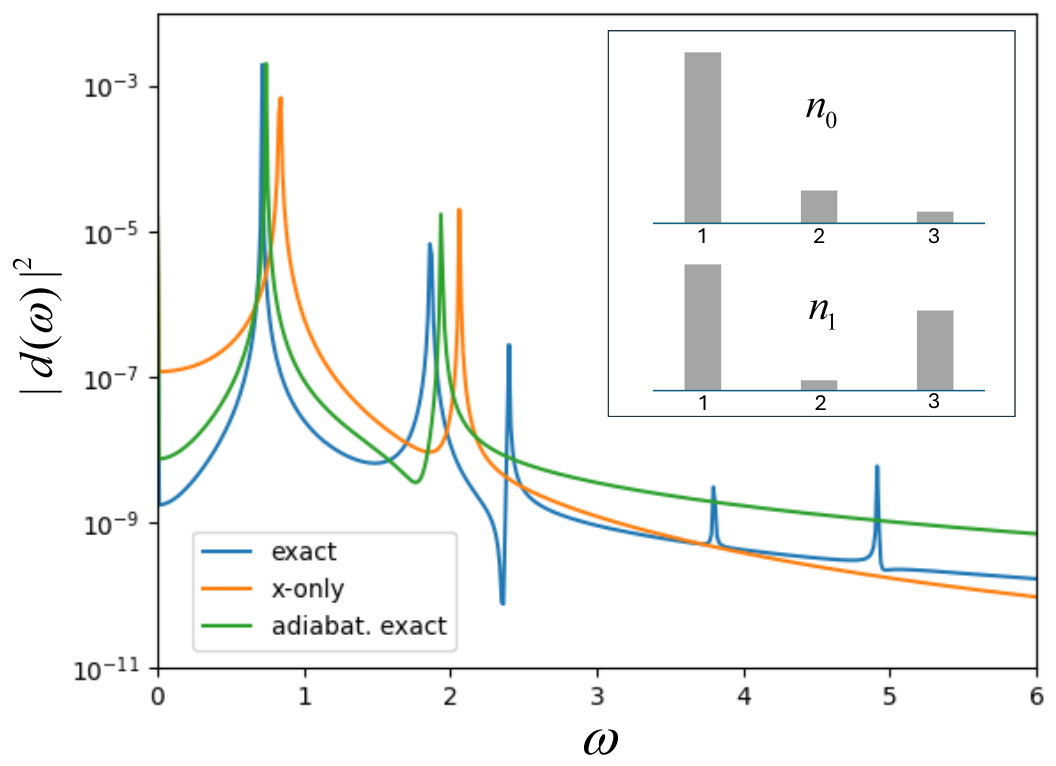}
  \caption{\label{fig2} Dipole power spectrum of two interacting electrons on a Hubbard trimer. The inset shows the site occupation
  of the ground state $(n_0)$ and first excited state $(n_1)$. The TDKS spectra with exchange-only and adiabatically exact approximation
  only reproduce the two single excitations but fail to capture the higher-lying double excitations.}
\end{figure}

As an example to illustrate some of the challenges in RT-TDDFT, let us consider a 3-point Hubbard lattice with two electrons in a singlet state.
Hubbard-type systems have become the favorite toy models in the (TD)DFT community due to their simplicity yet physical relevance.\cite{Capelle2013,Carrascal2015,Carrascal2018,Hill2023}
Here (using dimensionless units), we choose a linear 3-point lattice with scalar potentials $(V_1,V_2,V_3)=(-2,0,-0.5)$
and Hubbard model parameters $T=0.5$ (hopping) and $U=1$ (on-site interaction). The Hamiltonian of the model is given by
\begin{eqnarray}\label{Hubbard}
\hat H &=& -T\sum_{\sigma} (\hat c^\dagger_{1\sigma}\hat c_{2\sigma} + \hat c^\dagger_{2\sigma}\hat c_{3\sigma} + h.c.) + U\sum_{l=1}^3
\hat c^\dagger_{l\ua}\hat c_{l\ua}\hat c^\dagger_{l\da}\hat c_{l\da}  \nonumber\\
&+& \sum_{l=1}^3\sum_{\sigma}V_l \hat c^\dagger_{l\sigma}\hat c_{l\sigma} \:,
\end{eqnarray}
where $h.c.$ stands for Hermitian conjugate and $\hat c_{l\sigma}^\dagger$ and $\hat c_{l\sigma}$ are creation and annihilation operators
for electrons with spin $\sigma$ on site $l$. The system is asymmetric (see inset in Fig. \ref{fig2}),
with a ground state whose population is predominantly concentrated on point 1 and a first excited state where more population is found on point 3,
which means that we can use this system to illustrate charge-transfer processes. We solve the interacting 2-body Schr\"odinger equation numerically exactly, and the
associated TDKS system using the exact exchange approximation as well as the adiabatically exact xc potential.

Figure \ref{fig2} shows the dipole power spectrum $|d(\omega)|^2$, obtained using time propagation from the ground state triggered by a short and weak electric field kick,
and subsequent Fourier transformation of the time-dependent dipole moment $d(t)$. The first two peaks, around $\omega=0.8$ and $\omega=2$,
have predominantly the character of single excitations (with respect to a basis of exact Kohn-Sham eigenstates) and are thus well reproduced
by TDDFT, where the adiabatically exact xc potential is almost on top of the exact result. However, the three higher peaks are not captured by adiabatic TDDFT:
these have predominantly doubly excited character, which requires a memory-dependent xc potential.

\begin{figure}
  \includegraphics[width=0.95\linewidth]{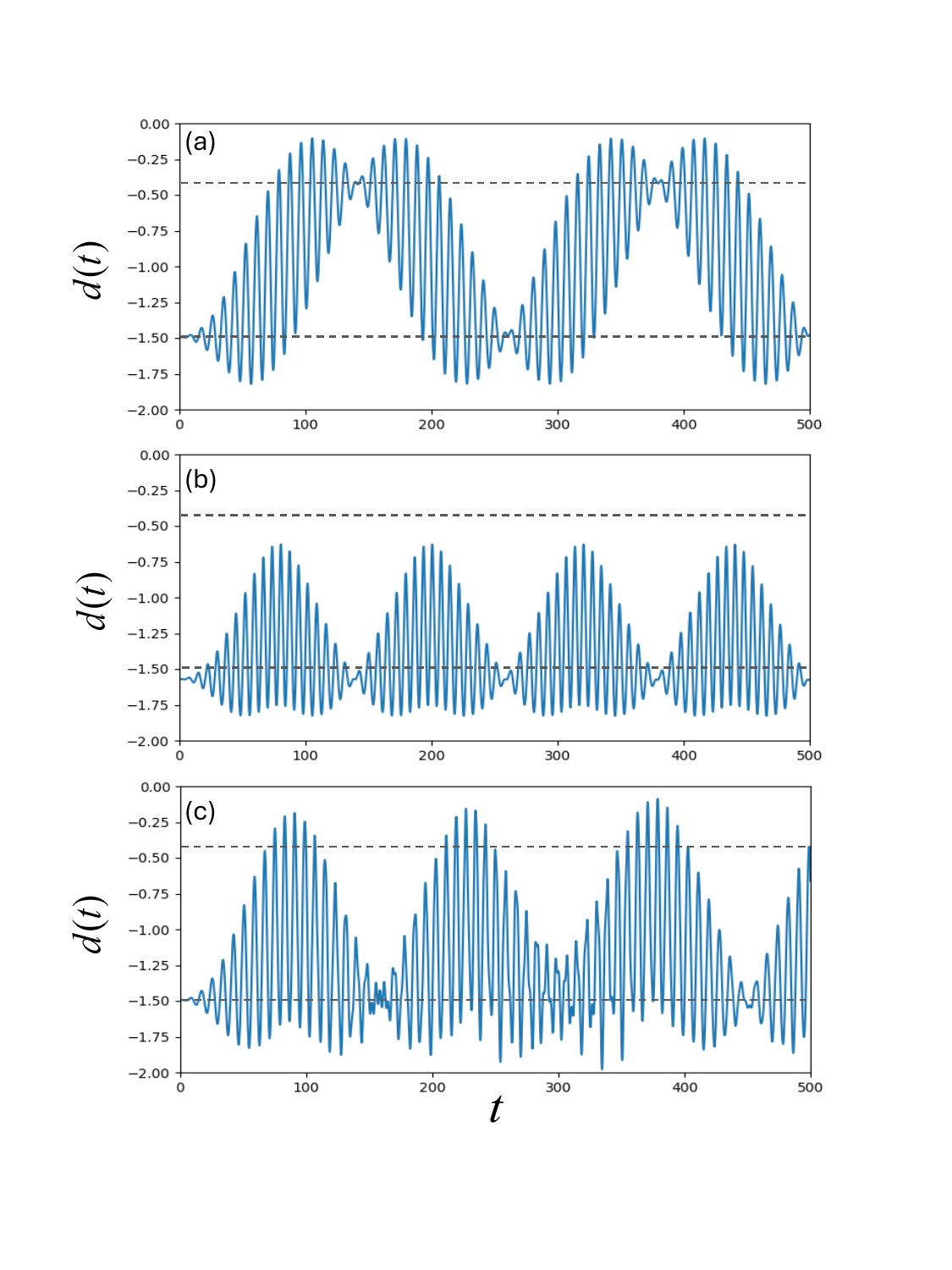}
  \caption{\label{fig3} Time-dependent dipole moment $d(t)$ for the Hubbard trimer, driven at the first resonance. (a) Exact solution of the 2-electron
  Schr\"odinger equation. (b) TDKS with exchange-only approximation. (c) TDKS with the adiabatically exact xc potential.
  The horizontal dashed lines indicate the exact dipole moments of the ground state $(-1.49)$ and the first excited state $(-0.43)$.}
\end{figure}

One might thus expect that the adiabatic xc functionals should give at least a good account of dynamical phenomena that predominantly involve single
excitations. To see whether this is the case, we consider so-called Rabi oscillations,\cite{Sakurai} where a periodic driving force at resonance with
the energy splitting in a two-level system induces periodic transfer of population between the ground and excited state.  Here,
we drive the Hubbard trimer at the first resonance, which should induce Rabi oscillations between the
ground and first excited state, whose dipole moments are $-1.49$ and $-0.43$, respectively. The time-dependent dipole moments of the driven system are shown in  Fig. \ref{fig3}.

The exact solution of the time-dependent 2-electron Schr\"odinger equation (top panel) clearly shows that the system exhibits Rabi oscillations, where the
dipole moment periodically switches between that of the ground state and that of the first excited state. On the other hand, the adiabatic TDDFT approximations fail completely:
the system clearly never fully transitions into the first excited state.

The failure of adiabatic xc functionals to describe Rabi oscillations was first noted over a decade and a half ago and subsequently explored in great detail.\cite{Ruggenthaler2009,Helbig2011,Fuks2011,Elliott2012,Fuks2013,Habenicht2014,Provorse2015,Oliveira2015} It was also noted that adiabatic xc functionals
can fail in achieving coherent control goals.\cite{Raghunathan2011,Raghunathan2012a,Raghunathan2012b}
The physical reason for this failure can be explained as follows.
Assuming we have the adiabatically exact xc potential, or a good adiabatic approximation, then the laser will initially be perfectly on resonance
with the excitation under consideration; energy will be absorbed very efficiently, the system gets shaken up and responds with strong dipole oscillations.
Very soon, however, the excited system falls out of resonance: the xc potential evaluated with the strongly excited density
(or site occupation, for lattice systems) leads to a shift (or detuning) of the
resonance with respect to the laser field. Hence, energy absorption becomes less efficient, and the system never quite reaches a point where all of the
population gets transferred to the upper level---in fact, it rarely makes it even a fraction of the way. The exact, memory-dependent xc potential, on the other hand,
is smart enough to keep the system in resonance with the laser field, regardless of the fraction of excited-state population.
The invariance of the resonance positions, a key property in time-resolved spectroscopies, is one of the more recent additions to the list of known exact conditions of
TDDFT.\cite{Fuks2015,Luo2016}

It should be mentioned that the failures of the adiabatic approximation noted above have been mainly discussed for rather small systems. Ranka and Isborn
studied the size dependence of such errors and found that the shortcomings of the adiabatic approximation seem to be much less pronounced
for larger molecules and materials.\cite{Ranka2023}

\begin{figure*}
  \includegraphics[width=0.95\linewidth]{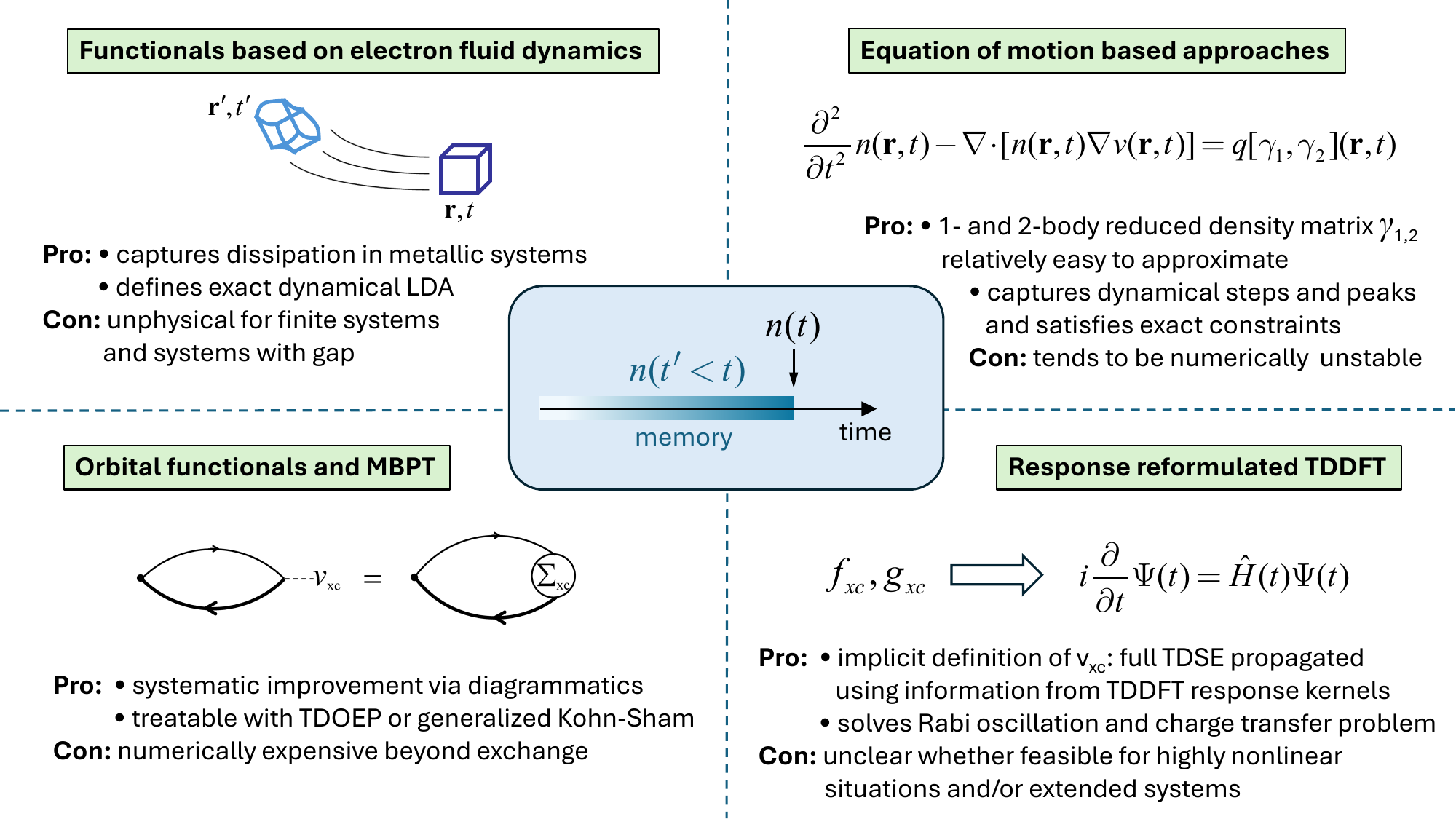}
  \caption{\label{fig4} Schematic summary of nonadiabatic TDDFT methods and their pros and cons. See text for more details and references.}
\end{figure*}

\subsection{Developments of nonadiabatic xc functionals}

One of the most difficult problems in TDDFT is nonadiabaticity. In linear response, this causes the xc kernel $f_{\rm xc}(\bfr,\bfr',\omega)$ to
be dependent on the frequency $\omega$, and in RT-TDDFT this causes $v_{\rm xc}(\bfr,t)$ to carry a memory of earlier times $t'<t$. Assuming
that such xc functionals have been found, this immediately causes practical challenges:
\begin{itemize}
\item In LR-TDDFT, the $\omega$-dependence of $f_{\rm xc}$ means that the Casida equation, which we will discuss further in Sec. \ref{sec:LR}, becomes a
nonlinear eigenvalue problem, which significantly complicates its solution.

\item In RT-TDDFT, when solving the TDKS equation numerically via time-stepping algorithms, densities and perhaps also TDKS orbitals at all
previous times must be carried along, which significantly raises the computational effort.
There is some evidence for metallic systems that TDKS electrons appear to ``forget'', i.e., most of the xc memory effects tend to be relatively
short-ranged into the past.\cite{Wijewardane2005} Whether this is more generally the case, especially in systems with discrete, well-separated energy levels, remains an interesting open question.
\end{itemize}

The remaining problem is to find explicit nonadiabatic approximations for the xc kernel and xc potential, respectively.
There is a long history in TDDFT, starting soon after its inception, of attempts to construct nonadiabatic xc functionals:
a comprehensive account can be found in Ref. \onlinecite{Ullrich2012}, and a concise review was recently published by Lacombe and Maitra.\cite{Lacombe2023}
A schematic summary of the various approximations is given in Fig. \ref{fig4}. We mention that there are nonadiabatic TDDFT approaches specifically
designed for transport through strongly correlated lattice model systems of quantum junctions, which are not further discussed here (see Refs. \onlinecite{Kurth2018,Covito2020}).

\subsubsection{Electron-liquid based functionals}

The earliest nonadiabatic TDDFT approaches arose from attempts to extend the LDA into the dynamical regime.\cite{Gross1985}
However, it was soon realized that nonlocality in time and nonlocality in space are inextricably connected to each other:
a small volume element of the electronic density distribution, located at space-time point $(\bfr,t)$, may have been at a
different position $\bfr'$ at an earlier time $t'<t$, due to the flow of currents in the system. It was therefore a natural
consequence that nonadiabatic TDDFT approximations were formulated using the language
of hydrodynamics,\cite{Vignale1996,Vignale1997,Dobson1997,Kurzweil2004,Tokatly2007,Thiele2009,Entwistle2020}
where xc effects enter in the form of viscoelastic stresses. The resulting formalism was successfully applied to describe dissipation
in collective excitations in metallic systems.\cite{Ullrich2001,Ullrich2002,Wijewardane2005} However, it turned out that it led to unphysical results for atoms and molecules,\cite{Ullrich2004} and it performed poorly for insulating solids.\cite{Berger2007}

\subsubsection{Orbital functionals}

Another way to construct nonadiabatic approximations is as explicit functionals of the Kohn-Sham orbitals. The first such approximation
was the time-dependent optimized effective potential (TDOEP) method.\cite{Ullrich1995,Goerling1997}
In the TDOEP scheme, the xc potential $v_{\rm xc}(\bfr,t)$ follows from an integral equation of the general form
\begin{equation}
\int_{-\infty}^t dt' \int d\bfr' v_{\rm xc}(\bfr',t') Q(\bfr,t,\bfr',t') = R(\bfr,t),
\end{equation}
where the integral kernel $Q(\bfr,t,\bfr',t')$ and the right-hand side $R(\bfr,t)$ depend on the occupied and unoccupied Kohn-Sham orbitals at all times $t'\le t$.
The TDOEP equation has been solved in the exchange-only limit, where the memory effects lead to frequency renormalizations.\cite{Wijewardane2008,Liao2017}

Beyond exchange, dynamical correlation effects can be treated by merging TDDFT with many-body perturbation theory, leading to
a diagrammatic construction of $v_{\rm xc}(\bfr,t)$ on the Keldysh contour, featuring nonequilibrium Green's functions and the self-energy $\Sigma_{\rm xc}$
(for details, see Ref. \onlinecite{Ullrich2012}). While of formal interest, this approach has so far not found practical applications in RT-TDDFT.
Instead, it turned out to be more fruitful to pursue approaches in which the nonequilibrium Green's functions are directly time propagated (via the Kadanoff-Baym equations
or time-dependent GW), rather than using them to construct a $v_{\rm xc}(\bfr,t)$.\cite{Attaccalite2011,Perfetto2015,Chan2021,Chan2023,Jiang2021,Perfetto2022,Reeves2024}

On the other hand, in LR-TDDFT, the diagrammatic approach has led to the construction of the so-called nanoquanta kernel:\cite{Reining2002,Onida2002}
The idea is to construct $f_{\rm xc}$ by re-casting the Bethe-Salpeter equation (BSE) in a TDDFT framework. The resulting xc kernel reproduces the
results of the full BSE,\cite{Adragna2003,Marini2003,Sottile2003,Sottile2007,Bruneval2005,Bruneval2006,Gatti2007,Gatti2011},
but at a rather high computational cost. However, it has led to important conceptual insights, and has also given rise to further approximations
such as the long-range corrected (LRC) kernel, which will be further discussed in Sec. \ref{sec:excitons}.

\subsubsection{Equation-of-motion based functionals}\label{sec:IID3}

A very different approach to the construction of nonadiabatic xc functionals starts from the equation of motion (\ref{EOM}).
A corresponding equation of motion can be written down for the TDKS system, and from this one obtains the following explicit expression
for the Hartree-xc potential, dropping the arguments $(\bfr,t)$ for brevity:
\begin{equation}
\nabla \cdot [n \nabla v_{\rm Hxc}]
=\nabla \cdot \left[ \bfF^{\rm kin} - \bfF^{\rm kin}_{\rm s}+ \bfF^{\rm int} \right].
\end{equation}
We have
\begin{eqnarray}
\nabla \cdot \left[ \bfF^{\rm kin} - \bfF^{\rm kin}_{\rm s}\right]
&=&
\frac{1}{2}\sum_{\mu\nu}\nabla_\mu \nabla_\nu(\nabla_\mu\nabla'_\nu - \nabla_\nu\nabla'_\mu)\nonumber\\
&&
\times [\gamma_1(\bfr,\bfr',t) - \gamma_{\rm 1s}(\bfr,\bfr',t)]\bigg|_{\bfr = \bfr'}
\end{eqnarray}
and
\begin{equation}
\nabla \cdot \bfF^{\rm int}
= 2\nabla\cdot \int d\bfr' \gamma_2(\bfr,\bfr',t)\nabla\frac{1}{|\bfr-\bfr'|} \:,
\end{equation}
where $\gamma_1$ and $\gamma_{1s}$ are the reduced 1-body density matrices of the interacting system and the TDKS system, respectively,
and $\gamma_2$ is the diagonal two-body density matrix of the interacting system. Thus, approximations for $\gamma_{1,2}$ immediately
lead to expressions for $v_{\rm xc}$. Such approximations were proposed and studied by Lacombe and Maitra,\cite{Lacombe2019,Lacombe2020,Lacombe2023}
and found to satisfy many important exact conditions and to reproduce key features of time-dependent xc functionals such as peaks and steps.
However, the numerical behavior was unsatisfactory as it led to instabilities.

\subsubsection{Response reformulated TDDFT}

Recently, Dar {\em et al.}\cite{Dar2024} proposed an interesting approach to circumvent the difficulties discussed above. The idea is to return to the time-dependent
many-body Schr\"odinger equation (\ref{SE}) and to expand $\Psi(t)$ in terms of many-body eigenstates, $\Psi(t) = \sum_n C_n(t) \Psi_n$.
Assume, for simplicity, that we start at $t_0$ from the ground state, and let $v(\bfr,t) = v_0(\bfr) + v_1(\bfr,t)$, where the time-dependent potential $v_1$ is switched on at $t_0$.
This then leads to an equation of motion for the expansion coefficients,
\begin{equation}\label{Ceq}
i \frac{d}{dt} C_m(t) = E_m C_m + \sum_n V_{mn}(t) C_n(t) \:,
\end{equation}
where the $E_m$ are the many-body energies and $V_{mn}(t) = \int d\bfr v_1(\bfr,t) \rho_{mn}(\bfr)$, where $\rho_{mn}(\bfr)$ are the transition densities
between the eigenstates $m$ and $n$. The equation of motion (\ref{Ceq}) is then propagated in time for $t>t_0$, with the initial condition
$C_m(t_0) = \delta_{m,0}$ (other initial conditions are also possible). The time-dependent density follows as
\begin{equation}\label{Cden}
n(\bfr,t) = \sum_{mn}C_n^*(t) C_m(t) \rho_{nm}(\bfr)\:.
\end{equation}
Notice that the time-dependent many-body wave function $\Psi(t)$ and the static many-body states $\Psi_n$ are never explicitly needed,
only the initial values of the expansion coefficients, $C_n(t_0)$, must be specified. The $E_m$ can be determined, in principle
exactly, from linear-response TDDFT, and so can the $\rho_{n0}$ (the transition densities between the ground and $n$th excited state). To determine
the transition densities between excited states, $\rho_{mn}(\bfr)$, $m,n>0$, second-order response TDDFT is needed.\cite{Parker2018}

\begin{figure}
  \includegraphics[width=\linewidth]{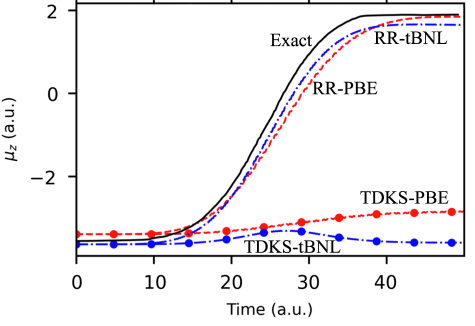}
  \caption{\label{fig5} Dipole moment $\mu_z$ associated with the resonantly driven charge transfer in the LiCN molecule, calculated using TDKS and RR-TDDFT.
  LiCN, a linear molecule, has a dipole moment in the degenerate second and third excited states opposite to that of the ground state, and a
  laser pulse at the excitation frequency should induce a large change in $\mu_z$ (see curve labeled ``Exact'').
  The adiabatic PBE \cite{Perdew1996} and tBNL \cite{Baer2005,Livshits2007} functionals fail with TDKS, but achieve complete dipole switching
  within the RR scheme. Adapted with permission from APS from Ref. \onlinecite{Dar2024}, \copyright 2024.}
\end{figure}

The TDKS scheme, Eqs. (\ref{TDKS}) and (\ref{TDKS_n}), and the response-reformulated (RR-TDDFT) scheme by Dar {\em et al.}, Eqs. (\ref{Ceq}) and (\ref{Cden}),
are both in principle exact. The difference lies in the way in which the xc functionals enter. TDKS evaluates $v_{\rm xc}$ with the full
time-dependent density, which can be very far from the ground state. RR-TDDFT, on the other hand, only requires xc functionals to be evaluated
close to the ground-state density, even for strongly nonlinear situations; this is much more favorable when only adiabatic approximations are available.
Indeed, this seems to solve the Rabi oscillation problem, at least for simple systems.\cite{Dar2024} More generally, the RR-TDDFT approach
could be a valuable alternative to TDKS in situations where significant amounts of density are moved around during time propagation (see Fig. \ref{fig5} for another example,
resonantly driven charge transfer).

A final remark: RR-TDDFT seems to resemble the TDCI (time-dependent configuration interaction) approach, which is
well known in the literature.\cite{Peng2018,Liu2019,Mehmood2024} However, in TDCI, the expansion of the many-body wave function $\Psi(t)$ is in terms
of Slater determinants rather than many-body eigenstates. Therefore, RR-TDDFT should be more economical in terms of basis size.

\begin{table*}
\caption{\label{table1}List of codes with real-time TDDFT capabilities. }
\begin{ruledtabular}
\begin{tabular}{ccc}
Name & URL &  Implementation \\
\hline
\text{Octopus\cite{Octopus2020}}   & octopus-code.org   &  Real-space grid based, TDDFT-Maxwell and QED  \\
\text{ELK\cite{Dewhurst2016}}      & elk.sourceforge.io     &  Full-potential all-electron LAPW   \\
\text{Salmon\cite{Salmon2019}}     & salmon-tddft.jp        &        Real-space grid, norm-conserving pseudopotentials \\
\text{Exciting\cite{Exciting2021}} & exciting-code.org  & Full-potential all-electron LAPW\\
\text{PySCF\cite{PySCF2022}}       & pyscf.org &    Gaussian basis functions, written in Python\\
\text{MISTER-T}\cite{Chen2024}    & data.mendeley.com/datasets/psymy4ddnw/1  & Quantum optimal control using TDKS\\
\text{QRCODE}\cite{Choi2024}   &  www.bmwong-group.com/   & Massively parallelized TDDFT for periodic systems   \\
\text{NWChem}\cite{Lopata2011,Bruner2017,NWChem}  & nwchem-sw.org  & Gaussian basis functions or plane waves\\
\text{Siesta}\cite{SIESTA}         & siesta-project.org & Atomic-like basis functions \\
\text{Abinit}\cite{Abinit2025}     &   abinit.org &  Implementation in PAW basis \\
\text{CP2K}\cite{CP2K,Hanasaki2025} & cp2k.org   & Molecular dynamics,  plane-wave and Gaussian basis\\
\text{FHI-aims}\cite{Hekele2021,Abbott2025}    & fhi-aims.org  & Numerical atomic orbital basis\\
\text{Q-Chem}\cite{Qchem}           & q-chem.com & Gaussian basis functions\\
\text{MolGW}\cite{MolGW}           & molgw.org & Gaussian basis functions \\
\text{Qbox/Qball/INQ}\cite{Draeger2017,Shepard2021,INQ2021,Kononov2022}& gitlab.com/npneq/inq & Plane waves or real-space grids\\
\text{Turbomole}\cite{Turbomole,Muller2020}   & turbomole.org &  Gaussian basis sets\\
\text{GPAW}\cite{GPAW2024}         & gpaw.readthedocs.io & Python-based, using PAW basis\\
\text{TDAP}\cite{Ma2016,Guan2022}         & everest.iphy.ac.cn/tdap & Numerical atomic orbital basis sets \\
\text{RMG}\cite{Jakowski2025}                  & rmgdft.sourceforge.net  & Real space basis and pseudopotentials\\
\text{DFTB+}\cite{Xu2023}         & dftbplus.org  & DFT tight binding method\\
\text{QDD}\cite{Dinh2022,Dinh2024,Heraud2025}                 & data.mendeley.com/datasets/2fg47zm4mz/1  & Real-space grid, quantum dissipative dynamics\\
\text{Yambo}\cite{Yambo2009,Yambo2019}    & yambo-code.eu  & Interfaces with ground-state DFT codes\\
\text{Chronus Quantum}\cite{Williams-Young2020} & github.com/xsligroup/chronusq\_public  & General treatment of spin and relativistic effects
\end{tabular}
\end{ruledtabular}
\end{table*}

\subsection{Computational developments}

There have been many exciting new computational developments in TDDFT in the last few years, the majority of which are for RT-TDDFT.
Linear-response TDDFT (used mainly for calculating excitation energies) has been well-established for longer, and we will discuss its progress in more
detail in Section \ref{sec:LR} below. Here, we focus on the computational aspects of propagating the TDKS equation, discussing
advances in algorithms, implementations, and code development.

In the early 2000s to 2010s there were only very few codes available for RT-TDDFT, but since then there has been a tremendous amount of
progress in code development. Today, prospective users have a choice of over 20 RT-TDDFT capable codes with a large variety of
features and functionalities and techniques of implementation, many of them open source, see the list codes given in Table \ref{table1}. In addition to this, many
customized implementations of RT-TDDFT exist which have not been officially released. Some of these implementations interface with widely
used ground-state DFT software to generate the initial Kohn-Sham wave functions which are then time propagated; many codes also take advantage
of the Libxc library of xc functionals.\cite{Libxc2012,Libxc2018} Users and developers now have a
choice of many well-tested time propagation methods, which have been extensively reviewed in the literature.\cite{Castro2004,Schleife2012,Pueyo2018,Zhu2018}

Amongst the many algorithmic developments in (TD)DFT, let us highlight two of them. The first is the adaptively compressed exchange (ACE)
operator technique, \cite{Lin2016} which greatly reduces the computational cost of applying the nonlocal exchange operator, such as
in Hartree-Fock or in hybrid calculations. The trick is to replace the high-rank Fock operator
with a low-rank approximation that produces the same results if acting on the occupied orbitals.
The resulting ACE exchange operator has the form of a projector, with a
computational cost similar to that of a nonlocal pseudopotential. The ACE technique
has now been implemented in many of the standard electronic structure codes, and can also be used in hybrid-RT-TDDFT.
Other efficient implementations of exact exchange have also been reported.\cite{Liang2011,Shepard2024}

The second algorithmic development to be mentioned here is the parallel transport gauge,\cite{Lin2018,Jia2019} which significantly accelerates the time propagation
in RT-TDDFT by allowing the user to choose a much larger time step. To do this, a gauge transformation of the TDKS orbitals is
found so that the transformed orbitals vary as slowly as possible without changing the density matrix. A related technique, which also produces
significant speedups of the time propagation, is achieved by using a basis of adiabatic eigenstates.\cite{Chu2001,Son2009,Russakoff2016,Ghosal2022}

Another interesting approach to make (TD)DFT calculations more efficient is via stochastic methods, which takes advantage of the
sparseness of the resulting TDKS and exact exchange Hamiltonian matrix and hence scales well for large systems. Stochastic methods have been implemented in RT-
as well as LR-TDDFT.\cite{Vlcek2018,Sereda2024,Chen2025}

Lastly, no review of computational developments would be complete without mentioning machine learning (ML), which has
nowadays become one of the main hot areas in DFT.\cite{Bogojeski2020,Kalita2021,Chen2023,Kelley2024,Luise2025}
Applications of ML in TDDFT are now beginning to emerge at a rapid pace and at various levels:
to capture correlation effects in $v_{\rm xc}(\bfr,t)$, including memory, through training on time-dependent model data sets\cite{Suzuki2020,Bhat2022}
or, within the adiabatic approximation, via unsupervised learning;\cite{Yang2023}
to make large-scale calculations more efficient by using ML to explore large parameter spaces;\cite{Ward2024}
or to directly predict dynamical properties or optical spectra of electronic systems without actually doing any TDDFT
calculations, once the neural network is suitably trained.\cite{Ramakrishnan2015,Bhat2020,Bai2022,Bhat2024,Shapera2025,Boyer2025}

\section{Progress in linear-response TDDFT}\label{sec:LR}

The vast majority of applications of TDDFT is still being done in the linear-response regime. The key quantity of LR-TDDFT is the response function
$\chi_{\alpha \beta}(\bfr,\bfr',\omega)$ of the
many-body system, which describes the response of a physical observable $\alpha$, such as the density, magnetization, or current, to a corresponding perturbation
$\beta$, for instance a scalar potential, an electric or a magnetic field. Most applications use the density-density response function,
so we will drop the labels $\alpha,\beta$ in the following. The response function is obtained in principle exactly as
\begin{eqnarray}\label{Dyson}
\chi(\bfr,\bfr',\omega) &=& \chi_{\rm s}(\bfr,\bfr',\omega) +\int \! d\bfr_1 \! \int \! d\bfr_2 \, \chi_{\rm s}(\bfr,\bfr_1,\omega)\nonumber\\
&&
{}\times f_{\rm Hxc}(\bfr_1,\bfr_2,\omega) \chi(\bfr_2,\bfr',\omega) \:,
\end{eqnarray}
where $\chi_{\rm s}$ is the corresponding response function of the Kohn-Sham system, which is relatively easy to calculate. The Hartree-xc kernel is given by
$f_{\rm Hxc}(\bfr,\bfr',\omega) = |\bfr-\bfr'|^{-1} + f_{\rm xc}(\bfr,\bfr',\omega)$, where the xc kernel is defined as
\begin{equation}
f_{\rm xc}(\bfr,\bfr',\omega) = \int d(t-t') e^{i\omega(t-t')} \left.\frac{dv_{\rm xc}[n](\bfr,t)}{\delta n(\bfr',t')}\right|_{n_0(\bfr)} \:.
\end{equation}
In practice, $f_{\rm xc}$ needs to be approximated, which is often done using the adiabatic approximation, Eq. (\ref{fxc_adia}). Note that we here
consider the linear response with respect to the ground state; a nonequilibrium TDDFT response theory has been formulated in Refs. \onlinecite{Fuks2015,Luo2016}.

The Dyson-type equation (\ref{Dyson}) can be recast into an equation that yields the exact excitation energies $\Omega$ of the many-body system
as a (non-Hermitian) eigenvalue problem known as the Casida equation:\cite{Casida1995,Ullrich2012}
\begin{equation} \label{Casida}
\left( \begin{array}{cc} {\bf A} & {\bf K} \\ {\bf K} & {\bf A} \end{array} \right)
\left( \begin{array}{c} {\bf X} \\ {\bf Y} \end{array} \right)
= \Omega
\left( \begin{array}{cc} -{\bf 1} & {\bf 0} \\ {\bf 0} & {\bf 1} \end{array} \right)
\left( \begin{array}{c} {\bf X} \\ {\bf Y} \end{array} \right),
\end{equation}
where the matrix elements of $\bf A$ and $\bf K$ are given by
\begin{eqnarray}
A_{ia,i'a'}(\omega) &=& \delta _{ii'} \delta_{aa'} \omega_{ai} +
K_{ia,i'a'}(\omega)
\\
K_{ia,i'a'}(\omega) &=&\int d \bfr \int d\bfr' \varphi_{i}^*(\bfr) \varphi_{a}(\bfr)
\nonumber\\
&\times& f_{\rm Hxc}(\bfr,\bfr',\omega) \varphi_{i'}(\bfr')\varphi_{a'}^*(\bfr')
\hspace{5mm}
\end{eqnarray}
and $i,i'$ and $a,a'$ run over occupied and unoccupied Kohn-Sham orbitals, respectively, and $\omega_{ai} = \epsilon_a - \epsilon_i$
are Kohn-Sham excitation energies. From the eigenvectors $\bf X,Y$ the oscillator strengths and, hence, the absorption spectrum of
the system can be calculated.

Equation (\ref{Casida}) is implemented nowadays in practically all major DFT-based electronic structure codes. A simplification, known as the Tamm-Dancoff approximation, is obtained by neglecting the off-diagonal $\bf K$-terms. The resulting simpler equation ${\bf A} {\bf Y} = \Omega {\bf Y}$ is often preferred over
the full Casida equation, either to save computational cost or to avoid numerical instabilities.\cite{Herbert2022}

In practice, LR-TDDFT strongly depends on the choice of xc kernel. The vast majority of applications use the adiabatic approximation, see Eq. (\ref{fxc_adia}),
where the $\omega$-dependence of $f_{\rm xc}$ is ignored and a linearized ground-state xc functional is employed.
The performance of many of the common xc approximations to describe low-lying excitations has been assessed for large molecular test sets,\cite{Jacquemin2009,Jacquemin2010,Laurent2013} and it was found that standard hybrid functionals such as PBE0 and B3LYP perform
with reliable and useful accuracy, with mean errors of about 0.25 eV. Other properties such as excited-states geometries and vibrational
frequencies are reproduced with accuracies that are comparable to those of ground-state DFT. With the so-called optimally tuned range separated hybrid
functionals,\cite{Stein2010,Kronik2012} a reliable prediction of the color of molecular systems\cite{Hartstein2024} and of excitonic properties (see Sec. \ref{sec:excitons})
has become possible. 

LR-TDDFT for applications in chemistry has been the subject of several excellent recent reviews,\cite{Casida2012,TDDFT2022,Herbert2023,Knepp2025}
and we won't go into any further details here. Among the many highlights that are currently at the forefront of LR-TDDFT in (photo)chemistry,
the proper description of conical intersections (or, more generally, the topology of ground- and excited-state potential energy surfaces) remains one of the most
important topics;\cite{Taylor2023} it now appears that the method of choice for conical intersections is spin-flip TDDFT.\cite{Herbert2022,Lee2025}
Another noteworthy topic is TDDFT for the quadratic response.\cite{Zahariev2014,Parker2018,Zaidi2024}

\begin{figure}
  \includegraphics[width=0.9\linewidth]{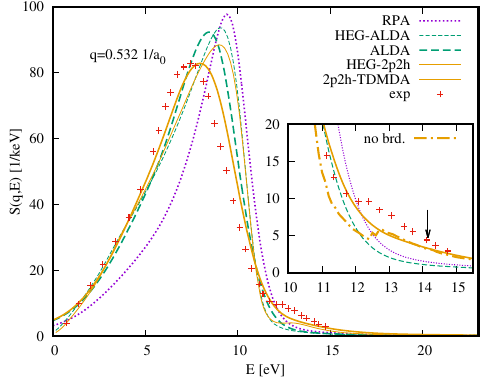}
  \caption{\label{fig6} Dynamic structure factor of sodium, at momentum $q=0.531/a_0$, comparing experiment with different levels of theory.
  The nonadiabatic TDDFT calculations (TDMDA) reproduces the two-plasmon excitation feature, indicated by the arrow in the inset. Reproduced with permission from APS from
  Ref. \onlinecite{Panholzer2018}, \copyright 2018.}
\end{figure}

\begin{figure*}
  \includegraphics[width=0.9\linewidth]{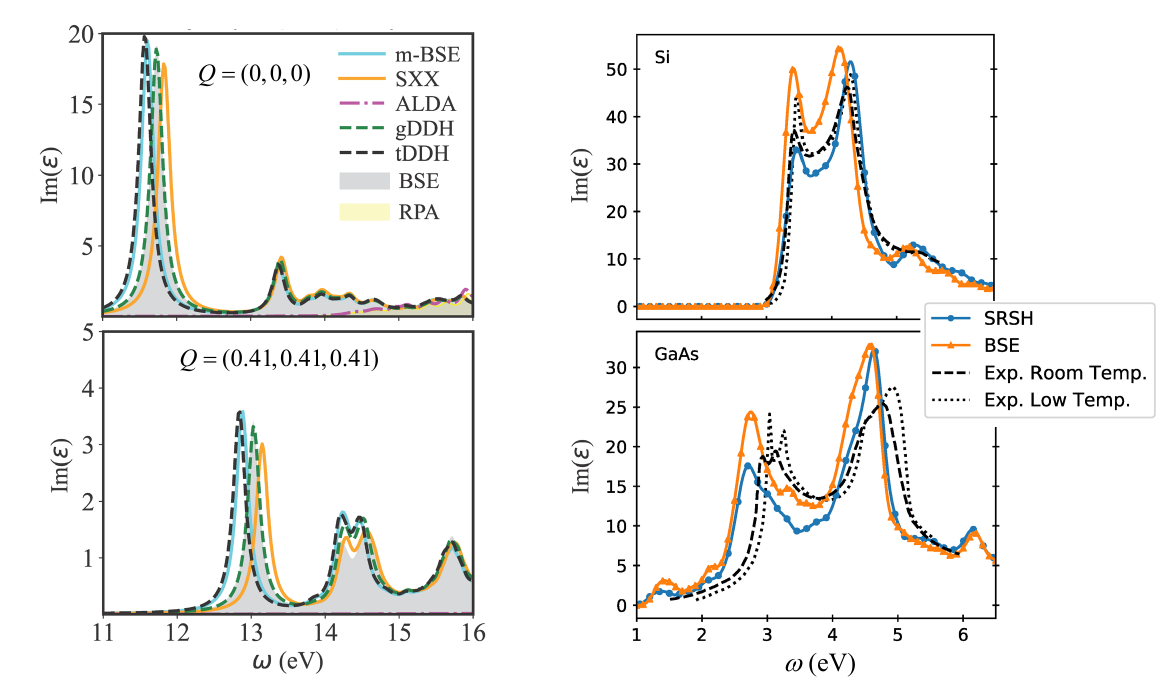}
  \caption{\label{fig7} Left panels: imaginary part of the dielectric function for LiF, for zero and finite momentum transfer $Q$. The best agreement
  with the BSE is obtained with the global DDH functional. Adapted with permission from APS from Ref. \onlinecite{Alam2025}, \copyright 2025.
  Right panels: optical absorption spectra for Si and GaAs, obtained with optimally tuned screened range-separated hybrid functionals.
  Adapted with permission from APS from Ref. \onlinecite{Wing2019}, \copyright 2019.}
\end{figure*}

\subsection{Nonadiabatic xc kernels}

The xc kernel is a strongly frequency-dependent quantity: this has been nicely illustrated for simple model systems in which $f_{\rm xc}$ can
be obtained exactly.\cite{Ruggenthaler2013,Thiele2014,Woods2021}
The frequency dependence of $f_{\rm xc}$ is essential to describe a variety of more challenging excitations:
double or multiple excitations,\cite{Maitra2022} charge-transfer excitations,\cite{Maitra2017} conical intersections,\cite{Taylor2023}
or strongly correlated systems.\cite{Turkowski2017b,Turkowski2022}
In the case of double excitations, the so-called ``dressed TDDFT'' approach\cite{Maitra2004} has been shown to work well for weakly correlated
systems where one (or several) single excitation(s) and one double excitation are close to each other and well separated from other excitations
in the spectrum.\cite{Cave2004,Huix2011,Authier2020} In this approach, $f_{\rm xc}(\omega)$ is explicitly constructed within a small many-body subspace.
Other attempts to obtain xc kernels for double excitations use constructions based on the many-body BSE.\cite{Romaniello2009,Zhang2013,Rebolini2016}
These approximations have so far only be tested for a few small systems; it was found that they
do produce double excitations with reasonable accuracy, but may also lead to spurious excitations.\cite{Authier2020}
A recent application to butadiene demonstrated that dressed TDDFT can successfully capture the crossing of potential-energy curves, which is
essential for photoinduced dynamics.\cite{Dar2025} 

On the other hand, there has been considerable recent effort in constructing wavevector- and frequency-dependent xc kernels $f_{\rm xc}(q,\omega)$
for the homogeneous electron gas.\cite{Panholzer2018,Ruzsinszky2000,LeBlanc2022,Kaplan2023b,Nazarov2024,Dornheim2024}
Such kernels can provide improved descriptions of collective excitations of the electron liquid, including lifetimes. An example is given in
Fig. \ref{fig6}, which shows that the nonadiabatic xc kernel not only gives an overall better description of the dynamic structure factor of sodium,
it also reproduces the signature of a two-plasmon excitation.

\subsection{Excitons and other response properties in solids}\label{sec:excitons}

One of the most significant challenges in TDDFT has been the accurate and efficient description of excitonic features
in the optical absorption spectra of insulators and semiconductors.\cite{Onida2002,Botti2007,Ullrich2015,Turkowski2017}
It was recognized early on that standard local and semilocal
xc functionals (LDA/GGA) fail to capture excitons in solids, producing spectra very similar to the random phase approximation (RPA).\cite{Reining2002}
This is explained by the failure of these functionals to satisfy the requirement that for systems with a gap the xc kernel
must behave as $\sim 1/|\bfr-\bfr'|$ for large separations between $\bfr$ and $\bfr'$.\cite{Ghosez1997}

Over the past two decades, the problem of producing excitons in solids with TDDFT has been intensely studied by many groups.
Early work focused on constructing an excitonic xc kernel
by reverse-engineering from many-body theory, leading to the ``nanoquanta'' kernel, which was accurate but computationally expensive
(more so than the BSE).\cite{Reining2002,Onida2002,Adragna2003,Marini2003,Sottile2003,Sottile2007,Bruneval2005,Bruneval2006,Gatti2007,Gatti2011}
A key feature of this many-body xc kernel was its correct long-range behavior, which then motivated the introduction of
the so-called long-range corrected (LRC) kernel:\cite{Reining2002,Botti2004}
\begin{equation}\label{LRC_kernel}
f_{\rm xc}^{\rm LRC}(\bfr,\bfr') = -\frac{\alpha}{4\pi} \frac{1}{|\bfr - \bfr'|} \:,
\end{equation}
where $\alpha$ is, in practice, treated as an adjustable para\-meter. The LRC kernel is the simplest TDDFT approach that
yields excitons\cite{Yang2012a} (not counting the so-called contact excitons\cite{Botti2007}), and as such is of great conceptual importance. In practice, however, its performance
is mixed, yielding reasonable absorption spectra for small-gap semiconductors, but unrealistic spectra for large-gap
insulators with strongly bound excitons. Generally, it is not possible with LRC to obtain accurate exciton binding energies
and spectral shapes at the same time; this can cause severe overestimation of the oscillator strength of strongly bound excitons such as in LiF or
Ar.\cite{Byun2017,Byun2020} This is also the case for other members of the family of LRC-type xc kernels.\cite{Sharma2011,Sharma2014,Rigamonti2015,Friedrich2017,Trevisanutto2013,Berger2015,Terentjev2018,Cavo2020}

\begin{figure*}
  \includegraphics[width=0.9\linewidth]{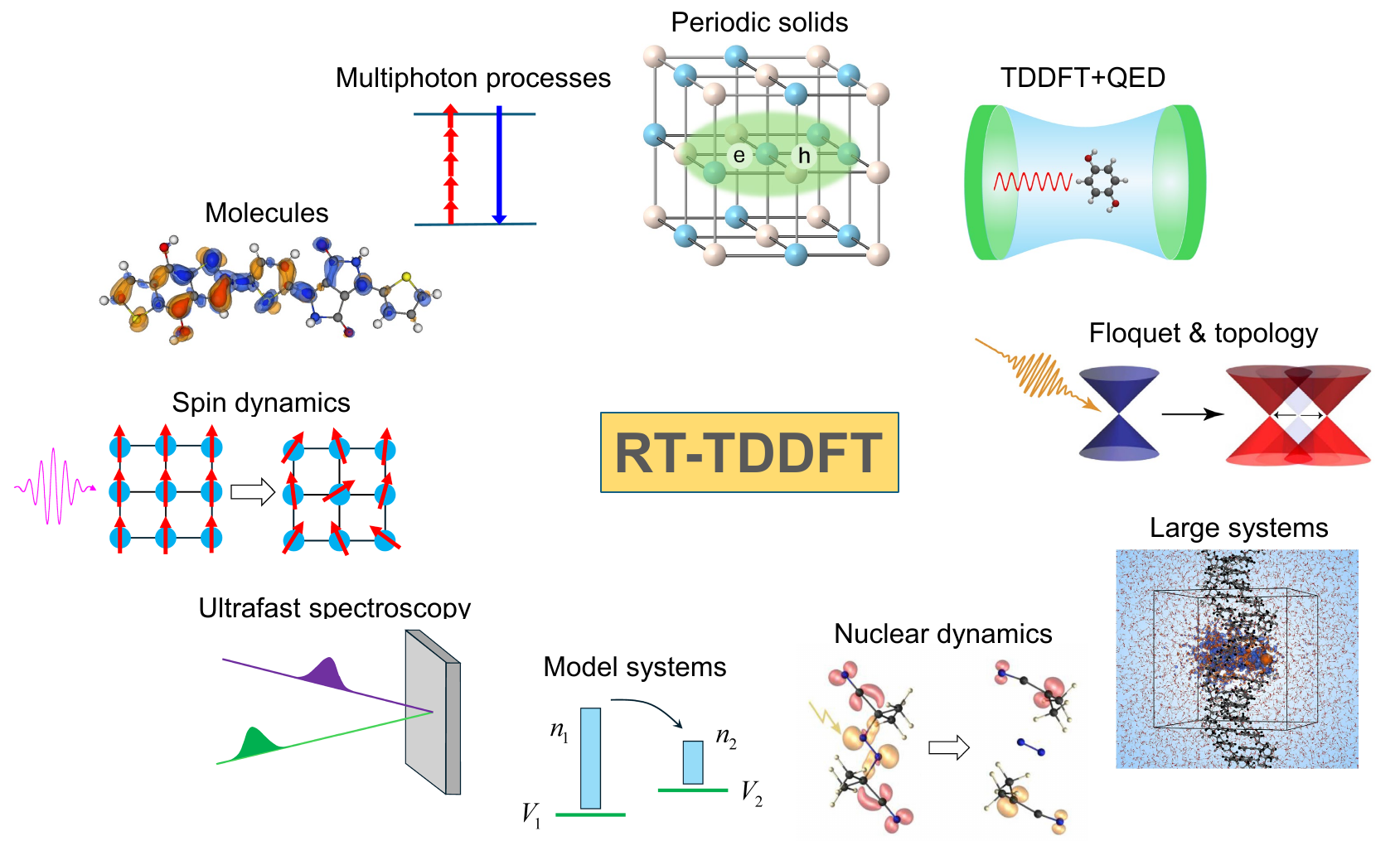}
  \caption{\label{fig8} Overview of applications of RT-TDDFT. Parts of this figure adapted with permission from Nature Portfolio from Ref. \onlinecite{Hubener2017}, \copyright 2017,
  with permission from APS from Ref. \onlinecite{Shepard2023}, \copyright 2023,   and with permission from ACS from Ref. \onlinecite{Kolesov2016}, \copyright 2016.}
\end{figure*}

Today, dielectrically screened hybrid functionals have become the most promising TDDFT approach for excitons in solids.
These types of hybrid functionals were found to be very successful in the calculation of band gaps of
solids,\cite{Shimazaki2010,Marques2011,Skone2014,Skone2016,Brawand2016,Brawand2017,Hinuma2017,Chen2018,Sagredo2025}
which motivated the development and implementation of screened exact exchange (SXX) and dielectric-dependent hybrid (DDH) functionals
for optical absorption spectra in solids.\cite{Yang2015,Sun2020a,Sun2020b,Alam2025}
These functionals give rise to xc coupling matrices of the form
\begin{equation}
K_{\rm xc}^{\rm DDH} = K^{\rm SXX} + (1-\gamma)K_{\rm xc}^{\rm ALDA} \:,
\end{equation}
which then formally enter the Casida equation (\ref{Casida}) (often within the Tamm-Dancoff approximation), or the Dyson equation (\ref{Dyson}).
Here, $K^{\rm SXX}$ is the exchange coupling matrix for the screened Coulomb interaction $\gamma/|\bfr - \bfr'|$, where
the screening parameter $\gamma$ is taken as the head of the inverse dielectric matrix, evaluated within RPA: $\gamma = (\varepsilon^{\rm RPA})^{-1}_{00}(\bfq = 0)$.
The left panels of Fig. \ref{fig7} show the imaginary part of the dielectric function of LiF for zero and finite momentum transfer $Q$,
calculated with different methods, but all using the same input band structure. The agreement between DDH and the BSE is nearly perfect, at a
fraction of the computational cost.\cite{Alam2025}

An alternative to the global hybrids discussed above are range-separated hybrids.
The first applications of optimally tuned range-separated hybrids for extended systems were for molecular solids.\cite{Refaely2015,Kronik2016,Manna2018}
More recently, dielectrically screened range-separated hybrid functionals were successfully applied to inorganic periodic solids;\cite{Tal2020}
if the range-separation parameter is determined via the so-called optimal tuning procedure, an additional step involving a transformation to Wannier orbitals is required.\cite{Wing2019,Ramasubramaniam2019,Gomez2024,Sagredo2024,Tal2020} As shown in the right panels of Fig. \ref{fig7} for Si and GaAs,
the agreement of this approach with the BSE and with experiment is excellent.

\section{New frontiers in real-time TDDFT}

The past decade has witnessed tremendous progress in the applications of RT-TDDFT to many different systems and effects, especially
in condensed-matter physics and materials science. A schematic overview illustrating the wide range of applications is given in Fig. \ref{fig8}.

\subsection{Dielectric response and excitonic effects in solids}\label{sec:RTexcitons}

RT-TDDFT has now become a widely available tool to simulate nonequilibrium processes in periodic solids from first principles.
The first such applications started 25 years ago with the pioneering work of Yabana and coworkers.\cite{Bertsch2000,Yabana2006,Yabana2012}
In this early work, a main goal was to calculate the optical properties of solids in the linear regime by triggering
induced current oscillations with a short electric-field ``kick'' associated with a suddenly switched on vector potential $\bfA(t) = \bfA_0 \theta(t-t_0)$,
and then propagating the TDKS equation of the form
\begin{eqnarray}\label{TDKSA}
\lefteqn{
i \frac{\partial}{\partial t} \varphi_j(\bfr,t)
=
\bigg[ \frac{1}{2}\left( \frac{\nabla}{i}+\bfA(t) + \bfA_{\rm xc}(t)\right)^2} \nonumber\\
&&{} + v(\bfr,t)
+v_{\rm H}(\bfr,t)  +v_{\rm xc}(\bfr,t)\bigg]\varphi_j(\bfr,t)\:.
\end{eqnarray}
Here, $\bfA_{\rm xc}(t)$ is an xc vector potential that will be specified below.
From the current density
\begin{equation}
\bfj(\bfr,t) = 2\Im\sum_j \varphi_j^*(\bfr,t) \nabla \varphi_j(\bfr,t) + n(\bfr,t)[\bfA(t) + \bfA_{\rm xc}(t)]
\end{equation}
one then obtains the macroscopic part
\begin{equation}
{\bf j}_{\rm mac}(t) = \frac{1}{\cal V} \int_{\rm cell} d\bfr {\bf j}(\bfr,t)
\end{equation}
($\cal V$ is the unit cell volume), which yields the frequency-dependent conductivity,
\begin{equation}\label{sigma}
\sigma(\omega) = -\frac{1}{A_0}\int_{t_0}^{t_0 +T} dt \, e^{i\omega t} f(t) j_{\rm mac}(t) \:,
\end{equation}
where $f(t)$ is a suitably chosen window function. From this the dielectric function follows as
\begin{equation}
\varepsilon_{\rm mac}(\omega) = 1 + \frac{4\pi i \sigma(\omega)}{\omega} \:.
\end{equation}
Alternatively, one can also calculate the dielectric function from the induced dipole moment.\cite{Sander2017}
The so obtained optical spectra are formally identical with absorption spectra from LR-TDDFT,\cite{Sander2017}
provided that the same level of approximate xc functional is used. In practice, the agreement depends of course on various
numerical issues such as time step, total length $T$ of the time propagation, algorithm, etc. It is also possible to include decoherence, at least
at a phenomenological level.\cite{Feng2022}

\begin{figure}
  \includegraphics[width=\linewidth]{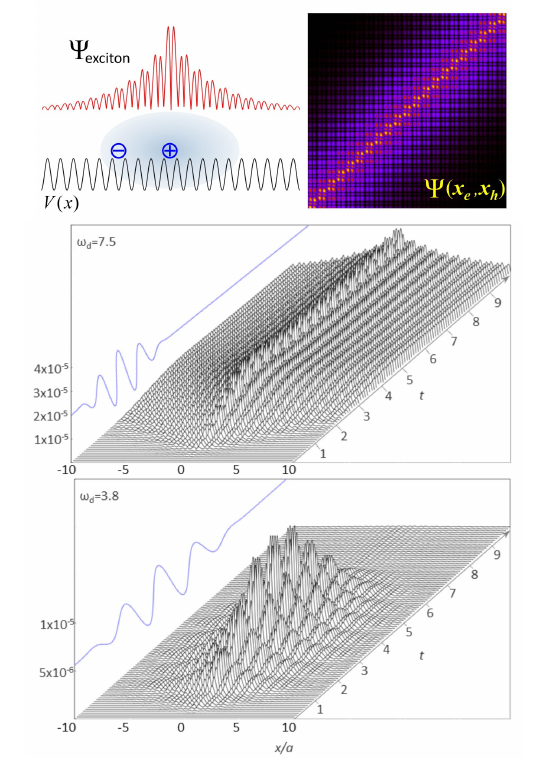}
  \caption{\label{fig9} Top left: exciton wave function $\Psi(x_e,x_h)$ for a 1D model solid, obtained from the Kohn-Sham TDM.
  Bottom panels: Time-dependent exciton wave function of the same model solid, following a pulsed excitation on resonance (middle)
  and below resonance (bottom) with the exciton. Adapted with permission from ACS from Ref. \onlinecite{Williams2021}, \copyright 2021.}
\end{figure}

The description of excitonic effects with RT-TDDFT was only accomplished very recently.\cite{Sun2021}
The idea is to construct a time-dependent LRC potential from the LRC xc kernel, Eq. (\ref{LRC_kernel}),
which only depends on the density response:
\begin{equation}
v_{\rm xc}^{\rm LRC}(\bfr,t) = -\frac{\alpha}{4\pi}\int d\bfr' \frac{\delta n(\bfr,t)}{|\bfr-\bfr'|} \:.
\end{equation}
Using a gauge transformation and the continuity equation, the long-range part of this -- which is the dominant contribution for excitonic effects -- can be transformed into an
xc vector potential:
\begin{equation}\label{LRC_A}
\frac{d^2}{dt^2} \bfA_{\rm xc}^{\rm LRC}(t) = \alpha \bfj_{\rm mac}(t) \:.
\end{equation}
This equation of motion for $ \bfA_{\rm xc}^{\rm LRC}(t)$ is solved together with the TDKS equation (\ref{TDKSA}).
The so-defined time-dependent LRC approach was successfully applied to small-gap semiconductors, but was found to be
unstable for systems with more strongly bound excitons.\cite{Sun2021} This stability issue was traced back to violations
of the zero-force theorem of TDDFT, which can be resolved by
incorporating additional terms into the equation of motion for $A_{\rm xc}^{\rm LRC}$.\cite{Dewhurst2025,Williams2025}
Thus, an RT-TDDFT approach to describe excitonic effects is now available.
We also mention that excitonic effects were studied with nonlocal exchange, using generalized RT-TDDFT implementations
within an LCAO basis\cite{Pemmaraju2018a,Pemmaraju2018b,Pemmaraju2020} and with real-space grids.\cite{Sato2015}

LR-TDDFT offers a variety of tools to visualize, characterize and analyze electronic excitations, including such quantities
as transition densities, attachment and detachment densities, natural transition orbitals, the transition density matrix (TDM),
or the particle-hole map.\cite{Tretiak2002,Tretiak2005,Bappler2014,Plasser2014,Mewes2015,Li2015,Li2016} Many of these tools are based on an analysis
of the Kohn-Sham orbitals or the Kohn-Sham density matrix. A recent review, mainly
from a chemistry perspective, was given by Herbert.\cite{Herbert2023b}

To visualize real-time dynamics it is of interest to extend these tools into the time domain. For the case of the TDM,
this is quite straightforward.\cite{Li2011,Williams2021} The time-dependent TDM of the Kohn-Sham system is defined as follows:
\begin{equation}\label{TDTDM}
\Gamma_{\rm s}(\bfr,\bfr',t) = \sum_{j=1}^N \left[ \varphi_j(\bfr,t)\varphi_j^*(\bfr',t) - \varphi_j^{\rm gs}(\bfr)\varphi_j^{\rm gs*}(\bfr')\right],
\end{equation}
i.e., the difference between the time-dependent and ground-state Kohn-Sham density matrices.
The time-dependent TDM can be used to visualize time-dependent exciton wave functions, including effects such as charge transfer or
field-induced exciton dissociation.\cite{Williams2021}
An example of the time-dependent exciton wave function is shown in Fig. \ref{fig9} for a one-dimensional model system.
For molecular systems, time-dependent natural transition orbitals were recently proposed.\cite{Zhou2021}

\subsection{Ultrafast and Nonlinear dynamics}

The true strength and versatility of RT-TDDFT emerges in the ultrafast, nonlinear regime,
which has seen tremendous experimental progress in recent years.\cite{Kruchinin2018,Caruso2025}
Let us now list some recent highlights from the perspective of condensed-matter physics and materials science.

\begin{figure*}
  \includegraphics[width=\linewidth]{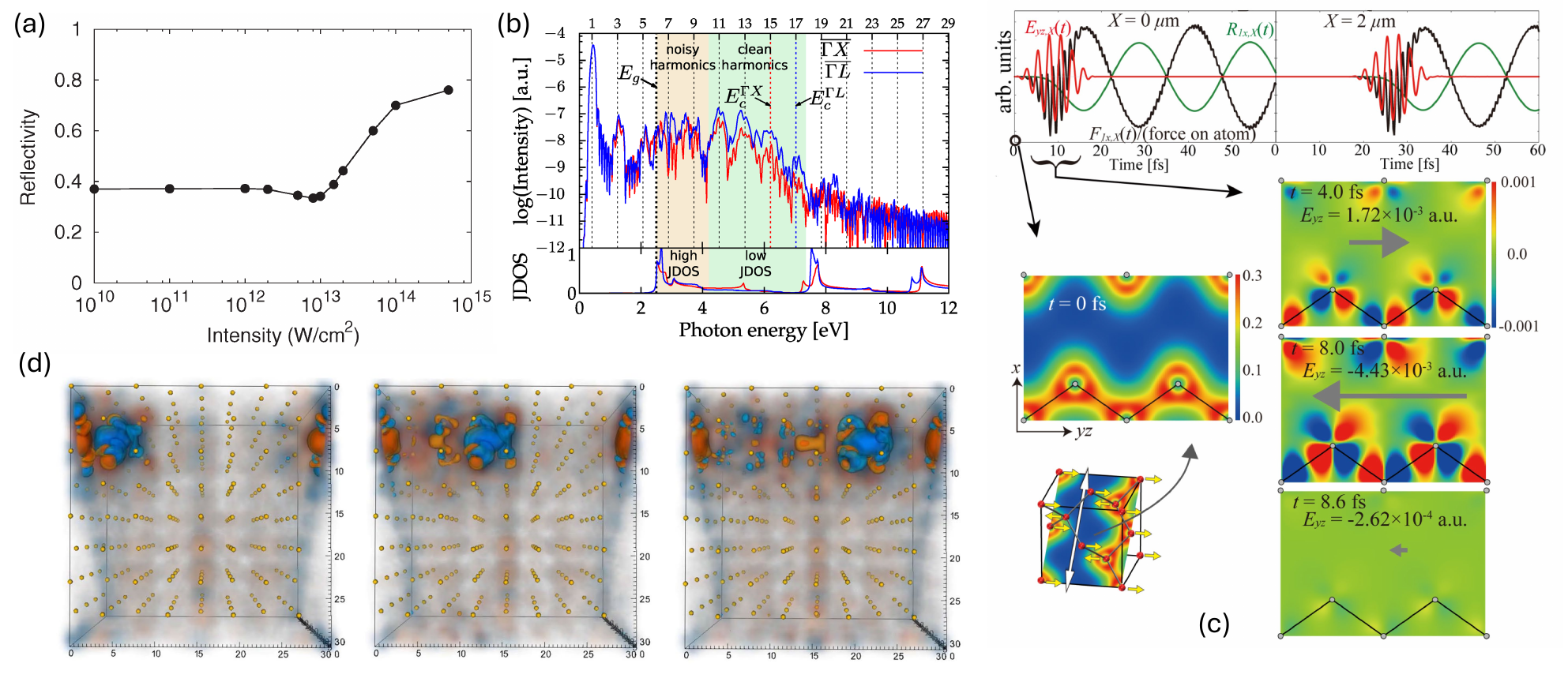}
  \caption{\label{fig10} Examples of applications of RT-TDDFT in solids.
  (a) Dielectric breakdown in bulk silicon. Reproduced with permission from APS from Ref. \onlinecite{Yabana2012}, \copyright 2012.
  (b) HHG in bulk silicon. Reproduced with permission from APS from Ref. \onlinecite{Tancogne2017a}, \copyright 2017.
  (c) Nonlinear response in diamond, including ionic forces. Reproduced with permission from APS from Ref. \onlinecite{Yamada2019}, \copyright 2019.
  (d) Snapshots of the density response in the wake of a proton passing through bulk Au. Figure courtesy of A. Schleife. }
\end{figure*}

\subsubsection{Nonlinear and ultrafast dielectric response}

Early implementations of RT-TDDFT for solids de\-monstrated the dielectric breakdown in insulators and semiconductors under intense irradiation:\cite{Otobe2008,Otobe2009,Yabana2012}
for instance, Fig. \ref{fig10}(a) shows how the reflectivity of bulk silicon drastically increases once the laser pulse intensity surpasses
a certain threshold, indicating the dominance of photoinduced free carriers. Subsequently, various studies addressed the
nonlinear response of the dielectric polarization in different bulk materials,\cite{Goncharov2013,Wachter2014,Su2017,Zhang2017}
as well as the electron and lattice dynamics of laser-excited graphene nanoribbons and 2D transition metal dichalcogenide heterojunctions.\cite{Lian2018}
Other studies focused on the generation of nonlinear photogalvanic currents\cite{Neufeld2021} or a detailed analysis of the population
of conduction band states to enable optically pumped lasing.\cite{Chen2025a}

The interaction of (classical) light and matter involves not only the quantum treatment of the electrons, but also
their coupling to the lattice degrees of freedom as well as the feedback to the propagating light pulses.
This challenge has stimulated the development of multiscale RT-TDDFT codes, coupled to Ehrenfest nuclear dynamics
and using Maxwell's equations to couple light and induced currents.\cite{Yamada2019,Hirokawa2022,Yamada2024,Tang2022}
In this way, large-scale simulations of light-driven effects such as coherent phonon generation,\cite{Shinohara2010,Shinohara2012,Kazempour2025}
or surface effects \cite{Sato2015a,Yamada2025,Chen2025b} have now become possible, see Fig. \ref{fig10}(c).

\subsubsection{High harmonic generation}

High-harmonic generation (HHG), a highly nonlinear phenomenon first observed in noble gases over four decades ago,
has now become firmly established as a widely studied experimental and theoretical field in solids and materials.\cite{Gruning2016,Ghimire2019,Yue2022}
RT-TDDFT has a long and successful history to describe HHG in atoms and
molecules,\cite{Ullrich1997b,Chu2001,Chu2001a,PenkaFowe2010,Chu2012,Wardlow2016,Coccia2022,Chu2024,Neufeld2024,Hamer2025}
and has now also emerged as an ideal tool for ab-initio simulation of HHG in solids.
First calculations focused on weakly correlated bulk materials such as $\alpha$-quartz,\cite{Otobe2016b} silicon [see Fig. \ref{fig10}(b)],\cite{Tancogne2017a,Tancogne2017b}
diamond,\cite{Floss2018} or 2D hBN.\cite{Tancogne2018b}
Other studies addressed HHG in liquid water\cite{Xu2025} and in strongly correlated NiO; \cite{Tancogne2018} the latter was made possible by
incorporating a (time-dependent) Hubbard $U$ in the RT-TDDFT calculation.\cite{Tancogne2017c}

A question of great interest is whether the spectral signatures of HHG can serve as optical probes of topological features in materials,
as suggested by studies on model systems. With the help of RT-TDDFT, a recent study\cite{Neufeld2023} concluded that HHG, unfortunately, tends to be an unreliable probe, and
that truly universal topological signatures in nonlinear optics appear to be unlikely.

\subsubsection{Electronic stopping power}

An important topic in materials science, highly relevant to characterize radiation damage in solids, is the calculation
of the stopping power. The stopping power -- which has the dimension of a force -- is defined as the loss of energy per unit length $dE/dx$
(or, equivalently, the amount of energy deposited into the host material) of a projectile along its path.
Earlier TDDFT studies calculated the stopping power of low-energetic projectiles using linear response theory,
including treatment of nonadiabatic viscoelastic effects at the level of the homogeneous electron liquid, see Ref. \onlinecite{Ullrich2012} for a summary.
Over the past decade, large-scale RT-TDDFT calculations have emerged as the state of the art of calculating the
electronic stopping power in a range of materials and at high projectile velocities,
where the dynamics is intrinsically nonadiabatic.\cite{Correa2012,Wang2015,Schleife2014,Schleife2015,Lim2016,Draeger2017,Caro2020,Shepard2023}
An example is shown in Fig. \ref{fig10}(d), which illustrates a proton travelling through gold along a channel,
leaving a highly disturbed electron density in its wake. Comparisons with experimental data generally show very good agreement
with the so calculated stopping powers.\cite{Schleife2015,Sand2019,Fu2020,Nunez2025}

RT-TDDFT has proved particularly important to calculate the stopping power of warm dense matter, in conditions associated with inertial confinement fusion;
a general overview and critical assessment of calculations across different computational platforms was given by Kononov {\em et al.}\cite{Kononov2024}
There have also been reports of the electronic stopping in liquid water, polymeric, and biological systems
calculated via RT-TDDFT.\cite{Yao2019,Tandiana2023,Shepard2023,Xu2024,Matias2024}

\subsubsection{Time-resolved spectroscopy}

\begin{figure*}
  \includegraphics[width=\linewidth]{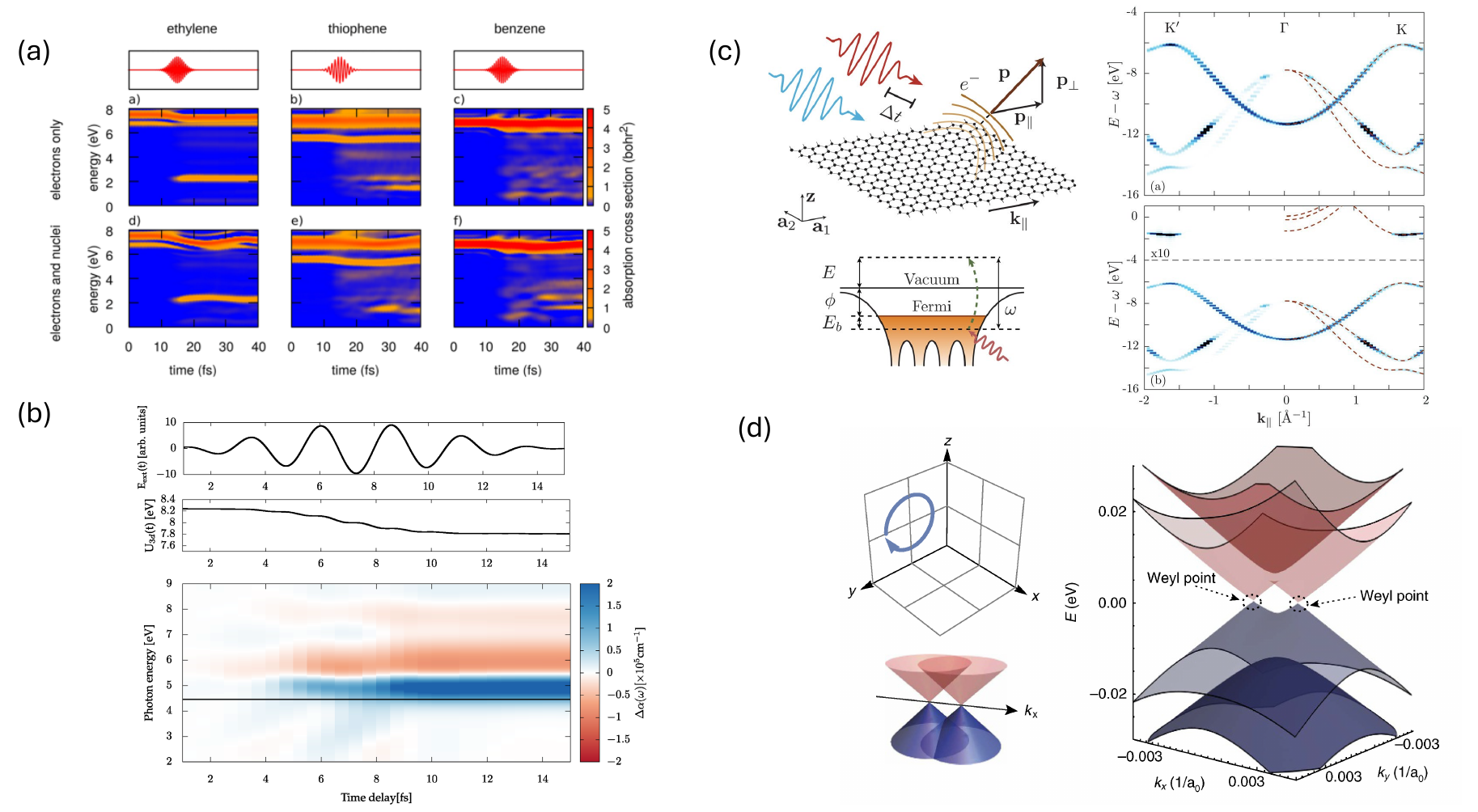}
  \caption{\label{fig11}  Examples of applications of TDDFT for time-resolved spectroscopy.
  (a) RT-TDDFT calculation of transient absorption spectra for ethylene, thiophene and benzene, with fixed (upper panels) and moving (lower panels) nuclei.
  Reproduced with permission from AIP from Ref. \onlinecite{Krumland2020}, \copyright 2020.
  (b) RT-TDDFT calculation of transient absorption of NiO: time-dependent electric field (top), evolution of on-site $U$ for the $3D$ orbitals of Ni (middle), and
  transient absorption spectrum (bottom). Reproduced with permission from APS from Ref. \onlinecite{Tancogne2020b}, \copyright 2020.
  (c) Schematic illustration of the time-resolved ARPES technique, and spectra for monolayer hBN probed in equilibrium (top right) and pumped (bottom right). Reproduced with
   permission from ACS from Ref. \onlinecite{Giovannini2017}, \copyright 2017.
  (d) Floquet engineering of Na$_3$Bi: circularly polarized femtosecond laser pulses switch the topological state from a Dirac point
  to two Weyl points. Reproduced with permission from Nature Portfolio from Ref. \onlinecite{Hubener2017}, \copyright 2017.
  }
\end{figure*}

A rapidly growing area of interest for RT-TDDFT is the simulation of time-resolved spectroscopy in a variety of systems and materials.
Experimental pump-probe techniques have been widely used ever since the availability of femtosecond laser systems, and
nowadays time-resolved ARPES (angle-resolved photoemission spectroscopy) has become a hot topic.\cite{Gedik2017,Boschini2024}
RT-TDDFT has been used to simulate transient absorption in molecules,\cite{Krumland2020,Moitra2023} where an excited-state
population is created by a pump pulse and then probed after a certain time delay, including effects caused by nuclear vibrations,
as illustrated for small organic molecules in Fig. \ref{fig11}(a).

Simulating transient absorption or other time-resolved spectroscopic
processes formally involves the time-dependent response function $\chi(\bfr,t,\bfr',t')$, which depends on $t$ and $t'$ individually,
rather than on $t-t'$, as would be the case if the response of a stationary state were considered.
This leads to a response formalism that is nonlocal in space and time, and challenging to implement and interpret.\cite{Perfetto2015}
RT-TDDFT is much more straightforward: transient absorption spectra can be obtained\cite{Sato2018} by calculating the induced currents
with the pump pulse alone, $\bfj_{\rm pump}(t)$, with pump and probe, $\bfj_{\rm pump+probe}(t)$, and then
evaluating the transient conductivity from the difference $\bfj_{\rm pump+probe}(t) - \bfj_{\rm pump}(t)$ in analogy with
Eq. (\ref{sigma}). This and related techniques have been successfully used to simulate transient absorption or the
dynamical Franz-Keldysh effect (i.e., light-induced modulation of the dielectric optical properties) for various
systems;\cite{Sato2014,Otobe2016a,Sato2018,Tancogne2020b} an example is shown in Fig. \ref{fig11}(b).
In particular, RT-TDDFT calculations have provided valuable support for experimental studies on the femtosecond population
transfer across the gap in Silicon\cite{Schultze2014} and attosecond transient absorption spectroscopy in diamond.\cite{Lucchini2016,Lucchini2020}

ARPES is an experimental technique to measure the electronic (quasiparticle) band structures of solids. Time-resolved ARPES
can show ultrafast changes of the electronic states, including replicas of bands (Floquet-Bloch states) generated by the oscillating electric fields.\cite{Gedik2017,Boschini2024}
The modelling of time-resolved ARPES using RT-TDDFT has been discussed in detail by De Giovannini and coworkers.\cite{Giovannini2017,Giovannini2022}
In essence, the outgoing electronic flux corresponding to photoionized electrons is calculated in a boundary region and then
analyzed to obtain the angle-resolved kinetic-energy spectrum. Fig. \ref{fig11}(c) illustrates the geometry used in the simulations, and shows
a comparison of time-resolved ARPES spectra of monolayer hBN from equilibrium and following a short pump pulse.

The concept of a time-dependent band structure is closely related to Floquet theory, which establishes the existence of quasi-stationary electronic states under
the influence of periodic driving fields. The connection between TDDFT and Floquet theory has been well explored in the literature,\cite{Maitra2002b,Maitra2007,Kapoor2013}
which makes it a highly promising approach for simulating Floquet engineering of quantum matter.\cite{Oka2019}
More recently, several applications of TDDFT to analyze and control Floquet topological phases in molecules\cite{Xu2024,Zhou2025} and
solids\cite{Hubener2017,Giovannini2020,Kazempour2021} have been reported. Notably, H{\"u}bener and coworkers demonstrated that circularly polarized femtosecond laser
pulses can be used to drive a 3D Dirac material, Na$_3$Bi, between topological states (Weyl semimetal, Dirac semimetal, and topological insulator),\cite{Hubener2017}
see Fig. \ref{fig11}(d).

\subsection{Applications in magnetism and spin dynamics}

\begin{figure}
  \includegraphics[width=0.9\linewidth]{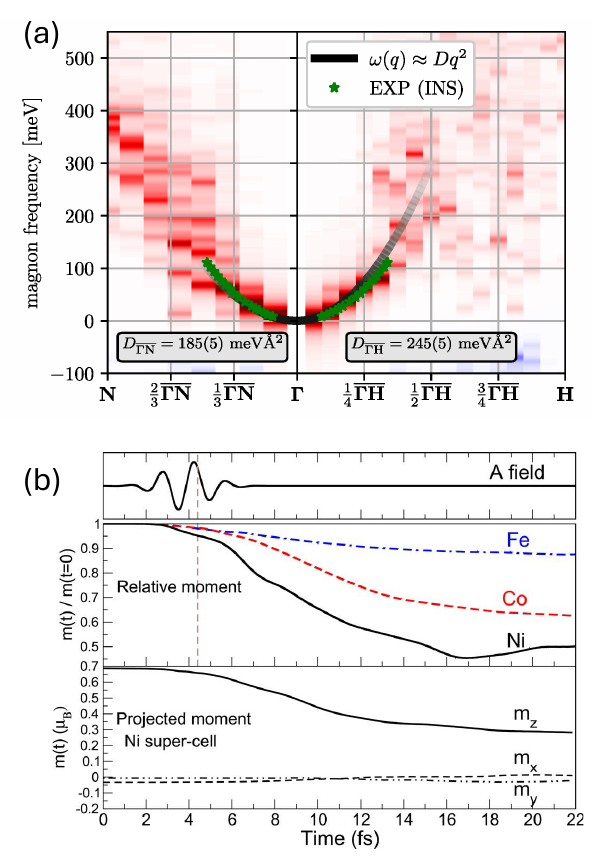}
  \caption{\label{fig12}
  (a) RT-TDDFT calculation of magnon dispersion in iron. Reproduced with permission from ACS from Ref. \onlinecite{Tancogne2020}, \copyright 2020.
  (b) Ultrafast demagnetization of ferromagnetic thin films, calculated with RT-TDDFT. Reproduced with permission from ACS from Ref. \onlinecite{Krieger2015}, \copyright 2015. }
\end{figure}

LR-TDDFT has been widely applied to calculate magnon dispersions in magnetic materials.\cite{Savrasov1998,Buczek2011,Cao2018,Skovhus2022,Zhang2025b,Binci2025}
Recent work has shown that magnons can also be described using a RT-TDDFT approach. The idea is, in principle, straightforward:
excite a spin fluctuation in a magnetic material with a short transverse magnetic field pulse, and from the
resulting time dependence of the magnetization ${\bf m}(\bfr,t)$ obtain the spin-wave frequencies via Fourier transformation.
This was practically realized in two different implementations: Tancogne-Dejean and coworkers, see Fig. \ref{fig12}(a), took advantage of a generalization of Bloch's theorem
with twisted boundary conditions to extract magnon frequencies at arbitrary wavevector;\cite{Tancogne2020}
Singh {\em et al.}, by contrast, used large supercells to select magnons with corresponding fixed wavevectors.\cite{Singh2020}

Beyond the linear regime, RT-TDDFT has found a broad range of applications in the study of nonlinear spin dynamics.
An important example is the so-called ultrafast demagnetization: a magnetic material or thin film, hit by a strong femtosecond
laser pulse, experiences a very rapid loss of magnetization.\cite{Beaurepaire1996} The mechanisms causing ultrafast demagnetization are quite complex and are still hotly
debated in the literature.\cite{Chen2025c} RT-TDDFT has made important contributions to the understanding of ultrafast demagnetization,
highlighting the crucial role of spin-orbit coupling in the interaction of the laser light and the magnetic moments of the material.\cite{Krieger2015,Elliott2016,Krieger2017}
Figure \ref{fig12}(b) shows how the magnetic moments of Fe, Co and Ni films are substantially reduced within a few femtoseconds following a pulsed
excitation. Other RT-TDDFT work on ultrafast demagnetization focused on various aspects,\cite{Zhang2009,Zhang2016,Zhang2018} including the impact of correlation\cite{Acharya2020,Barros2022,Mrudul2024} and the coupling to lattice dynamics.\cite{Wu2024,Zhou2024}

Another area in which RT-TDDFT plays an increasingly important role is in time-resolved
magnetooptics, which includes phenomena such as the magneto-optical Kerr effect (MOKE),\cite{Elhanoty2025}
transient magnetic circular dichroism,\cite{Dewhurst2020}
or generation of spin and valley current using circularly polarized light.\cite{Shallcross2022,Gill2025a,Gill2025b}
A major success for RT-TDDFT has been the prediction of the optically induced intersite spin transfer (OISTR) effect,\cite{Dewhurst2018,Elhanoty2022,Ryan2023}
which was subsequently experimentally observed\cite{Siegrist2019} and continues to be subject of joined experimental and theoretical studies.\cite{Moller2024,Geneaux2024,Gordes2025}
In addition to this, it has been shown that RT-TDDFT offers an elegant and efficient approach to obtain topological characteristics of materials
such as Berry curvature, Chern numbers and, hence, observable quantities such as anomalous and spin Hall conductivities or chirality-induced orbital and spin magnetism.\cite{Shin2019,Kim2023}

Apart from periodic solids, RT-TDDFT has found widespread applications to simulate spin dynamics in atoms, molecules, nanoparticles and model
systems.\cite{Stamenova2016,Simoni2017,Simoni2022,Roy2020,Li2024} Of particular interest here is the role of exchange-correlation torques,
which arise in (TD)DFT frameworks that explicitly treat noncollinear spin.\cite{Pluhar2019,Hill2023}

\subsection{Coupling to photon fields}\label{sec:4D}

A major current hot topic of TDDFT is the study of coupled electronic and photonic systems, which involves a quantum description of light.
The theory is based on the low-energy, nonrelativistic limit of quantum electrodynamics (QED), leading to the Pauli-Fierz Hamiltonian.
A corresponding QEDFT formalism was established by Ruggenthaler and coworkers,\cite{Ruggenthaler2011a,Ruggenthaler2014,Ruggenthaler2018}  see also earlier work
by Tokatly and Farzanehpour.\cite{Tokatly2013,Farzanehpour2014} The (TD)DFT formalism is then expressed in terms of the electronic and photonic
degrees of freedom, i.e., with current density and photonic displacement field (or, alternatively, current density and photonic vector potential) as basic variables, 
and approximate xc functionals can be derived.\cite{Pellegrini2015,Schafer2021,Flick2022,Tasci2025}

\begin{figure*}
  \includegraphics[width=\linewidth]{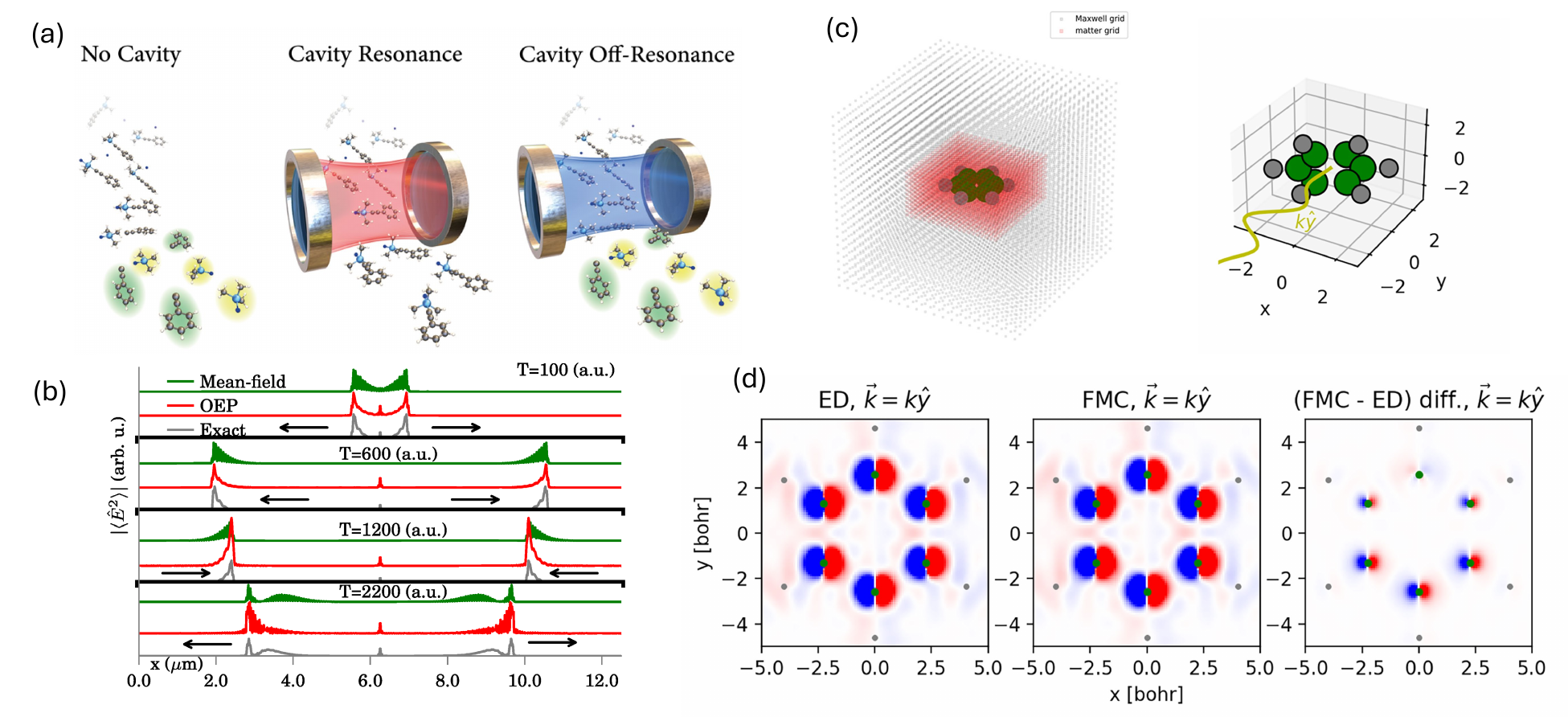}
  \caption{\label{fig13}  Applications of TDDFT to coupled electronic and photonic systems. (a) Illustration of cavity-mediated chemical reactivity, in which certain reaction outcomes can be selectively inhibited. Reproduced with permission
    from Nature Portfolio from Ref. \onlinecite{Schafer2022}, \copyright 2022.
  (b) Single-photon emission process in a cavity, comparing exact solution with mean-field and QEDFT-OEP solutions. Shown here is the time evolution
  of the intensity of the photon field. Reproduced with permission from NAS from Ref. \onlinecite{Flick2017}, \copyright 2017.
  (c) and (d) Coupled Maxwell-TDDFT calculations of benzene acted upon by a short laser pulse, illustrating retardation effects beyond the
    dipole approximation. Reproduced/adapted with permission from APS from Ref. \onlinecite{Bonafe2025}, \copyright 2025.
  }
\end{figure*}

The resulting formalism for light-matter interactions has a very wide range of applications, some of which are illustrated in Fig. \ref{fig13}.
A central interest is to study how the electronic structure and dynamics of matter is altered in a
photonic cavity, when photonic modes hybridize with electronic states.\cite{Flick2015,Flick2017,Flick2018,Flick2019,Malave2022,Sidler2022,Schafer2022}
For instance, Sch{\"a}fer {\em et al.} have investigated cavity-mediated chemical reactivity: as shown in Fig. \ref{fig13}(a),
strong resonant (red shaded) coupling between an optical cavity and vibrational modes can selectively inhibit a chemical reaction,
thus preventing the appearance of products (shaded green and yellow) present under off-resonant (blue shaded) conditions
or outside the cavity environment.

QEDFT also allows the study of photonic effects by themselves:\cite{Flick2017,Bakkestuen2025} as an example, Fig. \ref{fig13}(b) shows
the time evolution of a single-photon emission process, comparing exact solutions with mean-field and DFT solutions
using an approximation based on the OEP (optimized effective potential) method.\cite{Flick2017}

Very recently, the full minimal coupling Maxwell-TDDFT equations have been numerically implemented and used to study light-matter interaction
beyond the dipole interaction.\cite{Jestadt2019,Bonafe2025} This opens the way to study phenomena such as retardation effects during laser excitation, Cerenkov radiation of
electronic wave packets, or various types of new magneto-optical effects. Fig. \ref{fig13}(c) and (d) illustrate the interaction of a laser pulse
with a benzene molecule, showing substantial retardation as the pulse travels through the molecule.

\section{Concluding remarks}

In this paper, an attempt has been made to highlight the progress of TDDFT within the past decade.
We covered fundamental and formal aspects, algorithmic and code developments, and applications of TDDFT to many different types of systems
in the linear-response and real-time regimes. While not a comprehensive review, and perhaps a bit biased towards
applications in condensed-matter physics and materials science, TDDFT presents itself as a vibrant and rapidly evolving field with
a broad and diverse range of applications.

There are many exciting and important topics and examples of TDDFT that were not covered here:
atomic collisions;\cite{Kirchner2024} applications in plasmonics of metallic nanoparticles and nanostructures;\cite{Guidez2014,Varas2016,Herring2023}
orbital-free TDDFT;\cite{Jiang2021a,Zhang2025} time-dependent ensembles;\cite{Pribram2016,Daas2025} TDDFT for so-called strictly correlated systems;\cite{Cort2017,Cort2019}
or decoherence and dissipation effects.\cite{Dinh2018,Floss2019,Feng2022}

Another important topic beyond the scope of this article is the coupling of electron and nuclear dynamics. One of the
most significant recent developments in this context is the so-called exact factorization,\cite{Abedi2010,Abedi2012}
which proves that the full many-body wave function of the coupled electron-nuclear system can be written as
\begin{equation}\label{factorization}
\Psi(\underline{\underline{\bfr}},\underline{\underline{\bfR}},t) = \Phi_{\underline{\underline{\bfR}}}(\underline{\underline{\bfr}},t)
\chi(\underline{\underline{\bfR}},t)\:,
\end{equation}
where $\underline{\underline{\bfr}}$ and $\underline{\underline{\bfR}}$ are shorthand notations for all electronic and nuclear coordinates, respectively.
The decomposition (\ref{factorization}) is made unique and physically meaningful by the so-called partial normalization condition
$\int d\underline{\underline{\bfr}} \, |\Phi_{\underline{\underline{\bfR}}}(\underline{\underline{\bfr}},t)|^2=1$.
The so identified  electronic and nuclear wave functions, $\Phi$ and $\chi$, satisfy coupled equations of motion, including quantum effects such as the Berry phase.
This exact factorization formalism has led to numerous conceptually and practically important insights, such as an exact definition of time-dependent Born-Oppenheimer
surfaces and, hence, a new and rigorous framework of nonadiabatic molecular dynamics, as reviewed elsewhere.\cite{Agostini2021,Arribas2022}

To move the field of RT-TDDFT forward, a crucial task is the validation and critical assessment of functionals, methods and algorithms.
For instance, descriptions of pump-probe type spectroscopies using adiabatic approximations face the formal problem of shifting resonances discussed in
Sec. \ref{sec:Rabi}, see also the discussion in Ref. \onlinecite{Krumland2020}. However, quantitative assessments are difficult:
ultrafast time-resolved experiments in condensed-matter systems involve many additional complex factors,
such as electron-phonon interactions, defects and disorder, or effects related to sample and probe design; at present, these are beyond most RT-TDDFT implementations.
On the other hand, higher-level theories such as time-dependent GW tend to be too expensive for complex materials, and can only provide
benchmarks for relatively simple systems. These challenges need to be addressed for RT-TDDFT to become a quantitatively predictive tool for nonequilibrium dynamics in materials.

Thus, to conclude this perspective: the progress in TDDFT over the past decade has been truly remarkable, with no
indication of slowing down, and many interesting new research directions opening up.
The author hopes that this snapshot will be useful and inspiring to the current and future TDDFT research community.

\begin{acknowledgments}
Support from NSF Grant No. DMR-2149082 and from DOE Grant No. DE-SC0019109 is gratefully acknowledged.
\end{acknowledgments}

\section*{Data Availability}

The data that supports the findings of this study are available
within the article and on GitHub at: https://github.com/UllrichDFT/Rabi-trimer (Ref. \onlinecite{Rabi-trimer}).

\bibliography{TDDFT_refs}

\end{document}